\documentclass[a4paper,fleqn,usenatbib]{mnras}

\usepackage{newtxtext,newtxmath}

\usepackage[T1]{fontenc}
\usepackage{ae,aecompl}

\usepackage{graphicx}   
\usepackage{amsmath}    
\usepackage{soul}
\usepackage{tikz,xcolor,hyperref}

\definecolor{lime}{HTML}{A6CE39}
\DeclareRobustCommand{\orcidicon}{%
    \begin{tikzpicture}
    \draw[lime, fill=lime] (0,0)
    circle [radius=0.16]
    node[white] {{\fontfamily{qag}\selectfont \tiny ID}};
    \draw[white, fill=white] (-0.0625,0.095)
    circle [radius=0.007];
    \end{tikzpicture}
    \hspace{-2mm}
}

\newcommand{\orcidAM}{\href{https://orcid.org/0000-0002-9382-2542}{\orcidicon}}
\newcommand{\orcidWS}{\href{https://orcid.org/0000-0002-2393-8427}{\orcidicon}}
\newcommand{\orcidJDD}{\href{https://orcid.org/0000-0001-9704-6408}{\orcidicon}}

\newcommand{\mass}{\,M$_{\odot}$}   
\newcommand{\cpd}{\,d$^{-1}$}   
\newcommand{\dsct}{$\delta$\,Sct}   

\title[KIC\,10661783]{The eclipsing binary systems with $\delta$\,Scuti component \\
I. KIC\,10661783}

\author[A. Miszuda et al.]{
A. Miszuda$^{1}$\thanks{E-mail: miszuda@astro.uni.wroc.pl (AM)} \orcidAM,
W. Szewczuk$^{1}$\orcidWS,
J. Daszy\'nska-Daszkiewicz$^{1}$\orcidJDD
\\
$^{1}$Astronomical Institute, University of Wroc{\l}aw, ul. Kopernika 11, 51-622 Wroc{\l}aw, Poland\\
}

\date{Accepted XXX. Received YYY; in original form ZZZ}

\pubyear{2020}

\begin{document}
\label{firstpage}
\pagerange{\pageref{firstpage}--\pageref{lastpage}}
\maketitle

\begin{abstract}

We present a comprehensive study of the eclipsing binary system KIC\,10661783. The analysis of the whole \textit{Kepler} light curve, corrected for the binary effects, reveals a rich oscillation spectrum with 590 significant frequency peaks, 207 of which are independent. In addition to typical \dsct\ frequencies, we find small-amplitude signals in the low frequency range that, most probably, are a manifestation of gravity-mode pulsations. We perform binary-evolution computations for this system in order to find an acceptable model describing its current stage. Our models show that the binary KIC\,10661783 was formed by a rapid, almost conservative, mass transfer that heavily affected the evolution of both components in the past. One of the most important effects of binary evolution is the enormous enrichment of the outer layers of the main component with helium. This fact profoundly influences the pulsational properties of $\delta$\,Scuti star models. For the first time we demonstrate the effect of binary evolution on pulsational instability. We construct pulsational models of the main component in order to account for the mode instability of the observed frequencies. Whereas the single-star evolution model is pulsational stable in the whole frequency range, its binary-evolution counterpart has unstable modes, both, in high and low frequency range. However, to obtain instability in almost a whole range of the observed frequencies, the modification of the mean opacity at the depth corresponding to temperatures $\log T=4.69$\,K and $\log T=5.06$\,K was necessary.

\end{abstract}




 \section{Introduction}

Double-lined eclipsing binaries provide a unique opportunity to test various theoretical aspects of modern astrophysics. Their absolute parameters of masses and radii can be determined with outstanding accuracy, often with errors of  1\%  \citep{Torres2010}. Because of the precise estimates of stellar parameters, such systems are used as benchmarks for testing the theory of stellar evolution.
Of particular interest are binary systems with pulsating components as they can provide independent constraints on parameters of the model and theory.

The $\delta$\,Scuti ($\delta$\,Sct) stars are the intermediate mass stars, typically in the range of 1.5 - 2.5\mass, and with spectral types A0--F5 \citep[e.g.][]{Rodriguez2000,Aerts2010}.
They are located in the Hertzsprung-Russell (HR) diagram, in the lower part of the classical instability strip at the intersection with the main sequence \citep[e.g.][]{Dupret2005,Liakos2017}.
The pulsations of $\delta$ Sct variables are dominated by low-order pressure (p) and gravity (g) modes.
The $\delta$\,Sct instability region in the HR diagram partially overlaps with the $\gamma$\,Doradus ($\gamma$\,Dor) group, multi-periodic stars pulsating in high-order g modes with typical masses between 1.5 and 1.8\mass. They are located near the red edge of the classical instability strip of pulsations \citep[e.g.][]{Dupret2005,Aerts2010}.

Historically, $\delta$\,Sct and $\gamma$\,Dor stars constituted two separate groups. However, in the modern era of space-based photometric observations, hybrid $\delta$\,Sct/$\gamma$\,Dor pulsators with the frequencies typical for both types are becoming rather a rule than an exception \citep[e.g.][]{Grigahcene2010,Balona2015,Antoci2019}.

Low-order p/g modes in $\delta$\,Sct models are excited by the $\kappa$-$\gamma$ mechanism (the opacity mechanism) acting in the partial He\,II ionisation zone \citep[eg.][]{Pamyatnykh1999}. However as noted by \cite{Antoci2014}, the turbulent pressure in the hydrogen ionization zone can play a role in the excitation of high-order p modes.
 In the case of $\gamma$\,Dor it is widely believed that high-order g modes are excited by the interaction of convection and pulsations \cite[e.g.][]{Guzik2000,Grigahcene2005,Dupret2005,Xiong2016}.

A considerable fraction of \dsct\ stars are members of binary systems \citep{Liakos2017}. That gives an opportunity for a powerful test of stellar structure and evolution theory.
In particular, such systems enable the determination of current evolutionary status of components and the system age \citep[see e.g.,][]{Higl2017,Daszynska2019}, provide the possibility to study tidal interactions \citep{Bowman2019} and the effect of the mass exchange and its impact on the system evolution. As was shown by \cite{Claret2016,Claret2017,Claret2018,Claret2019}, double-lined eclipsing binary systems can provide also an estimate of overshooting efficiency from the convective core.

KIC\,10661783 is a close binary system of spectral type A5IV \citep{Frasca2016}. It was studied for the first time by \citet{Pigulski2009}, who reported that its light curve exhibits eclipses of both components. However, this study was based on the low number of \textit{ASAS} \citep{Pojmanski1997}  observational points and no detailed study of the system was possible.

The first thorough study of KIC\,10661783 was done by \cite{Southworth2011} who used short cadence (SC, Q2.3) and long cadence (LC, Q0-1) \textit{Kepler} satellite observations. The authors found 68 frequency peaks in the systems light curve with 58 identified as the independent frequencies. These independent peaks were assigned
to the primary component, defined as the more massive star at the current stage of the system evolution. From the modelling of the eclipsing light curve,
they derived two possible solutions. The first one with  detached geometry and the mass ratio $q=M_2/M_1=0.25$ and the second one with  semi-detached geometry of the system and $q=0.06$.
According to the authors, the second determination of $q$ was preferred by their preliminary spectroscopic measurements.
However, as the authors noted, their light curve fit was unsatisfactory and required unphysically high albedo for the primary star.
This mass-ratio discrepancy was resolved by \cite{Lehmann2013} who gathered 85 spectra of the system and determined the new value of $q=0.09$, that was slightly higher than the value for semi-detached configuration found by \cite{Southworth2011}.
The authors state that the system is a post-mass transfer detached binary with the fundamental stellar parameters: $M_A=2.100 \pm 0.028$\mass, $R_A=2.575 \pm 0.015\,R_{\odot}$ for the primary and $M_B=0.1913 \pm 0.0025$\mass, $R_B=1.124 \pm 0.019\,R_{\odot}$ for the secondary.
Recent work of \cite{Miszuda2020}, presented the preliminary analysis of the whole available \textit{Kepler} photometry of the system.

In this paper we present an extended study of KIC\,10661783 based on the \textit{Kepler} data.
In Section\,\ref{sec:observations} we give a short description of the used observations. Sections\,\ref{sec:binarymodelling} and \ref{sec:binaryevolution} are devoted to the eclipsing light curve modelling and binary evolution, respectively. In Section\,\ref{sec:freqanalysis} we analyse the variability of KIC\,10661783 and we extract the frequencies from its light curve residuals. Interpretation  of the oscillation spectrum is given  in Section\,\ref{sec:PulsationModelling}. Discussion and conclusions in Section\,\ref{sec:conclusions} end the paper. Finally, in Appendix\,\ref{sec:appendix}, we provide a list of all significant frequencies with their amplitudes and phases.

\section{Observations}
\label{sec:observations}

\textit{Kepler Space Telescope} was a space observatory under the subject of NASA space agency. Its core goal was to detect Earth-size planets orbiting solar-like stars, however it has a great potential for asteroseismology \cite[for the mission overwiew, see e.g.,][]{Borucki2010,Koch2010}. It was operating in its original mission form since early 2009 until mid-2013, resulting in $\sim$ 4 years of nearly continuous observations of the fixed field of view. After a technical failure the mission changed to \textit{K2} that observed various fields until the end of 2018, when the spacecraft ran out of fuel. The mission allowed for the creation of a unique catalogue of targets observed with unprecedented data quality, time span and duty cycle.
\textit{Kepler} was observing its target stars in two observational modes: short cadence (SC) and long cadence (LC). Each of the modes is composed of the summed up multiple 6.02\,s exposures followed by the 0.52\,s readout time, resulting in total 58.9\,s of exposure in short cadence and 29.4\,min exposure in long cadence \citep{Gilliland2010}.

KIC\,10661783 was observed almost continuously for over four years in the \textit{Kepler} long cadence mode. In addition, the star was observed in short cadence mode for over 2 years. These observations were reduced in a similar way as in \cite{Szewczuk2018}.
We extracted the flux from the target pixel files using the \texttt{PYKE} code \citep{Still2012} with the custom defined apertures (masks). Our masks contain pixels for which
signal to noise ratio exceeds 100. From this raw light curve we removed outliers. To this end the 4-$\sigma$ clipping was used. In order to remove some common instrumental trends we used the so-called co-trending basis vectors.
Then, some outliers were found and removed once again by an eye inspection. Finally, data were divided by second order polynomials fitted to the out-of-eclipses data in each quarter separately.

\begin{figure}
    \centering
    \includegraphics[width=0.5\textwidth,clip]{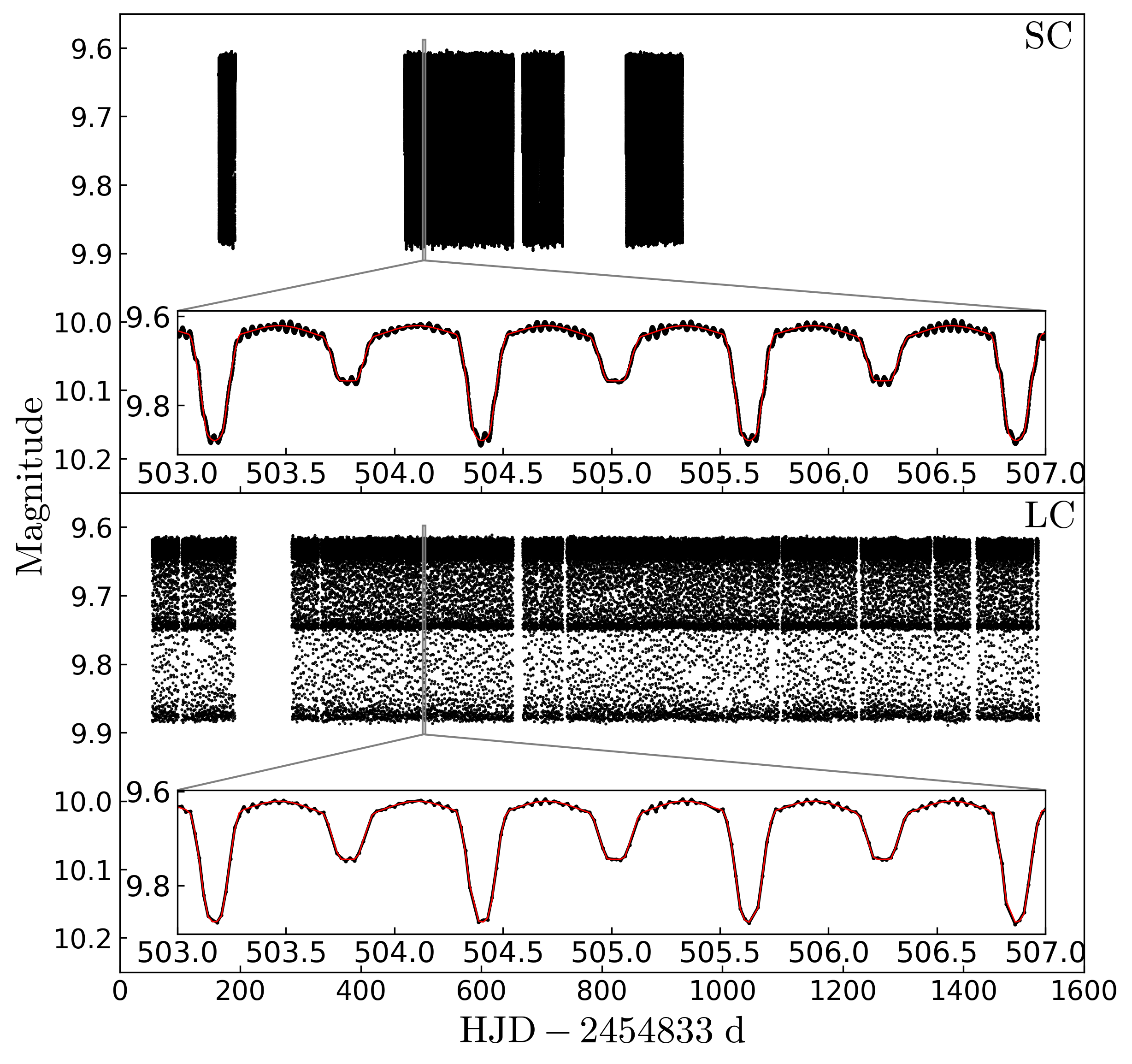}

    \caption{A comparison of the full \textit{Kepler} short cadence (top panel) and long cadence (bottom panel) light curves of KIC\,10661783. The insets show the zoomed area covering 4 days of observations in order to visualize both binary and pulsation variability. The \texttt{WD} model (described in Sect.\,\ref{sec:binarymodelling}) is marked with the red line.}
    \label{lc}
\end{figure}

The final light curve of KIC\,10661783 consists of over 58\,000 points spread over 1500 days for LC (Q0-Q17) and over 470\,000 points for SC (Q2.3, Q6.1-Q8.3, Q10.1-Q10.3) spread over a period of 769 days. A comparison of the light curves from both, the SC and LC modes can be seen in Fig.\,\ref{lc}. As one can see, the light curve exhibits both, the eclipses and pulsations.

\section{Binary light curve modelling}
\label{sec:binarymodelling}
We modelled the eclipsing light curve with the \texttt{JKTEBOP} code \citep{Southworth2004} and Wilson-Devinney code \citep[\texttt{WD}, see][]{Wilson1971,Wilson1979}. We used the \texttt{WD} version of May 22, 2015, in which the \textit{Kepler} passband is included, enabling us to properly model the passband-dependent features. The computed model for both SC and LC can be seen in the insets of Fig.\,\ref{lc}.

We started the modelling of the LC light curve with the \texttt{JKTEBOP} code. Firstly, we confirmed the results of \cite{Lehmann2013} that the system has a zero eccentricity. At this step we determined a rough value of the orbital period which was used as a starting value for the later analysis. The \texttt{WD} analysis was performed in a detached mode (mode 2) in a time domain.
Using the LC data we refined an orbital period value to $\rm P=1.231\,363\,26 \pm 0.000\,000\,03\,d$ which is slightly longer than the period derived by \citet{Southworth2011}, i.e., 
$1.23136220 \pm 0.00000024$\,d. The difference is lower than 0.1\,s, however exceeds the Southworth's 4$\sigma$ error.
We tested the possibility that the system may exhibit a period change, however we found no evidence for that. We also neglected the presence of third body as we noticed no signs of it in the light curve.

\begin{table}
    \centering
    \caption{The values of the effective temperature of the primary component of KIC\,10661783.}
    \label{tab:primary_temperature}
    \begin{tabular}{lc}

    \hline
    \hline
    Source & $T_{\rm eff}$  [K] \\
    \hline
    \cite{Lehmann2013} & 7764 $\pm$ 54 \\
    \citeauthor{GAIA2018} (\citeyear{GAIA2018}, DR2) & 7654 $\pm$ 286 \\
    \citeauthor{KIC2011} (\citeyear{KIC2011}, Kepler Input Catalog) & 7887 \\
    \hline
    \end{tabular}
\end{table}

The values of the primary's effective temperature gathered from the literature are summarized in Table\,\ref{tab:primary_temperature}.
We allowed for the wider range of the effective temperature derived by \cite{GAIA2018}, i.e., $T_{\rm eff}\in (7368, 7940)$\,K, for a better estimation of uncertainties in the parameters from the \texttt{WD} solution. This range includes the more precise determination of \cite{Lehmann2013}.
In the first run we fixed the effective temperature of the primary with the value resulting from the spectroscopic analysis of \cite{Lehmann2013}. The values of the semi-major axis and mass ratio were adopted from \cite{Lehmann2013} as well. We note, that the \texttt{WD} output of the effective temperature for the secondary fits well into its spectroscopic determination within 1$\sigma$ error. We were able to obtain a smaller value of albedo, i.e., 1.4 $\pm$ 0.14 comparing to 2.46 in \cite{Lehmann2013}, but it is still larger than 1.0.
Our study confirms some other results obtained by \citet{Lehmann2013}. In particular we obtained almost identical values of the inclination angle $i$, the surface potentials and the value of $T_{\rm eff}$ for the secondary.

\begin{figure*}
    \centering
    \includegraphics[width=0.8\textwidth,clip]{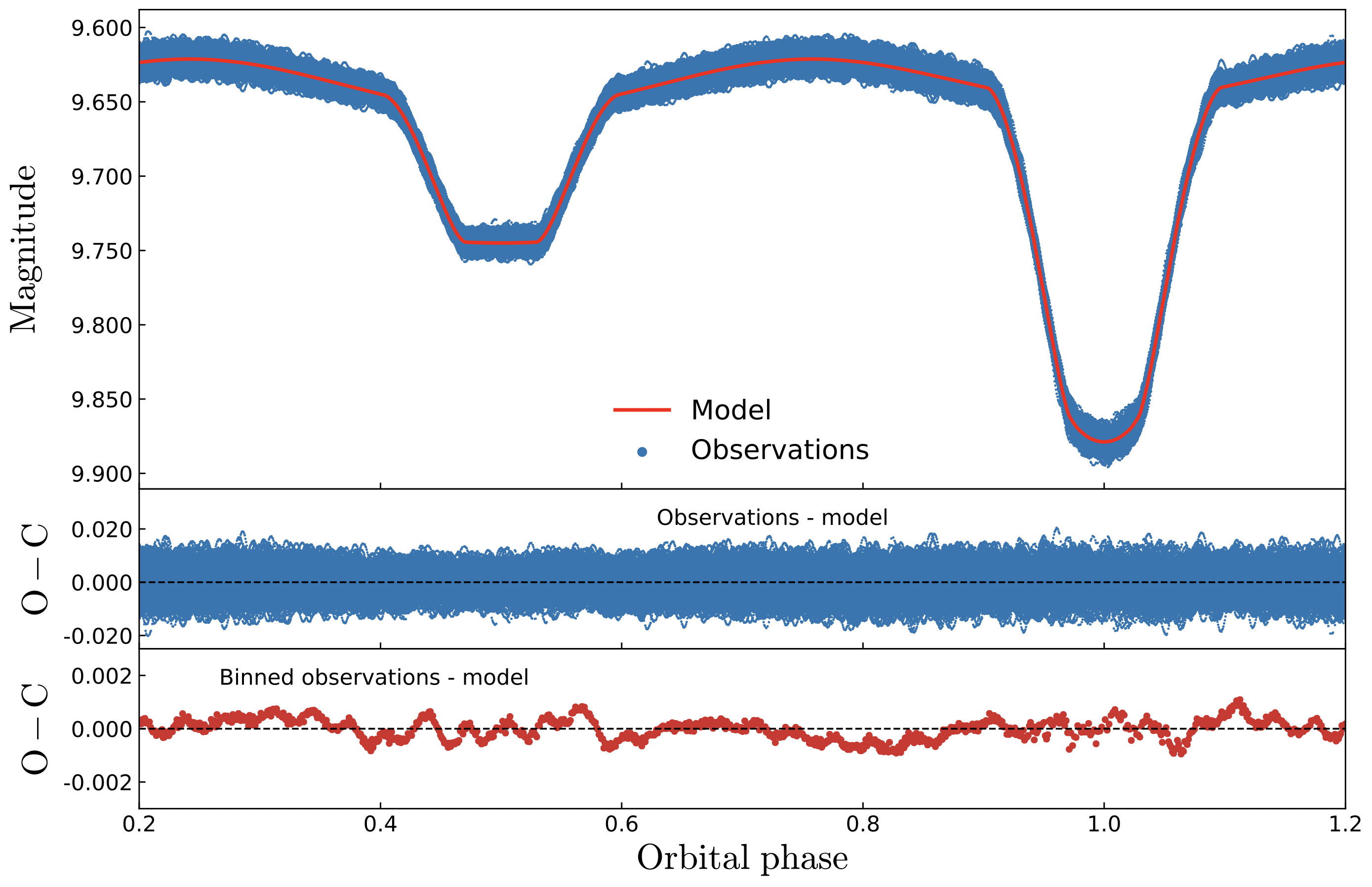}

    \caption{The upper panel presents the SC light curve phased with the orbital period. Those observations are depicted with blue points.
    The binary light curve model that was calculated using the \texttt{WD} code is plotted with the red solid line. The bottom panels presents the residuals after subtracting the binary light curve model from the observations and from the binned points.}
    \label{lc_phase_folded}
\end{figure*}

In the top panel of Fig.\,\ref{lc_phase_folded} with blue dots we plot all available SC observations as a function of the orbital phase. The \texttt{WD} binary light curve model is represented with a red solid line.
The middle and bottom panels show residuals from unbinned and binned observations respectively.
The SC observations were binned to 1000 points in a phase space. We do not plot those bins in the top panel as they coincide with the model almost perfectly.
However, the residuals from binned points exhibit regular trends. Since all pulsational variability, that is not a multiple of orbital frequency, has been averaged by phasing and binning the observations, this phenomenon is another kind of variability that manifests itself by the presence of the orbital harmonics.

To estimate uncertainties in all parameters of the system we checked the effect of changing the effective temperature of the primary on such parameters as:
$T_{\rm eff}$ of the secondary, luminosities and radii of both components etc.
To this aim we repeated the \texttt{WD} model fitting with the fixed minimum and maximum values of $T_{\rm eff}$. We also took into account the whole measured range of $q = 0.0898 - 0.092$ \citep{Lehmann2013}, different mesh sizes and two different limb darkening laws; linear and square root law.
The effect of changing the mesh size and the limb darkening law is negligible.
However, this time to speed up the computations, instead of using HJD times of observations, we supplied the program with the binned, phased-folded light curve.
Next, we determined the luminosities of both components, using a simple formula $\log L/L_{\odot} = 4 \log(T_{\rm eff}/T_{\rm eff \odot}) + 2 \log(R/R_{\odot})$ and adopting the 1$\sigma$ errors of both radius and temperature.
Table\,\ref{tab:system_parameters} gives a summary of all control and fixed parameters that were used during the \texttt{WD} runs.

\begin{table*}
\caption{The physical and orbital parameters of KIC\,10661783 found from the \texttt{WD} modelling. In the last two rows, we give the mesh size parameters.}
\label{tab:system_parameters}
\begin{tabular}{lccc}

\hline
\hline
Parameter                       & Primary               & Secondary         & System \\
\hline
Orbital period (days)       & ...                       & ...                       & $1.2313632588 \pm 3.26 \times 10^{-8} $ \\ 
$dP/dt$ (days/year)                     & ...                   & ...                       & $0.00 \pm 0.16\times 10^{-9}$ \\
Time of primary minimum (BJD-2\,454\,900) & ... & ...            & $164.5464621147 \pm 4.34 \times 10^{-5}$ \\
Orbital inclination (degree) & ...                   & ...                       & $82.03 \pm 0.01$ \\
Orbital eccentricity $e$      & ...                   & ...                       & $0.0^{\star}$ \\
Semi-major axis ($R_{\odot}$)& ...              & ...                       & $6.375^{\star1}$ \\
Mass ratio $q=M_2/M_1$  & ...                   & ...                       & $0.09109^{\star1}$ \\
Mass ($M_{\odot}$)            & $2.100^{1}$  & $0.1913^{1}$     & ... \\
Radius ($R_{\odot}$)          & $2.5793 \pm 0.0223$     & $1.1320 \pm 0.0020$           & ... \\
$T_{\rm eff}$ (K)                 & $7654 \pm 286^{\star}$& $6136 \pm 203$            & ... \\
$\log L/L_{\odot}$        & $1.3113 \pm 0.0652$     & $0.2121 \pm 0.0575$           & ... \\
Surface potential, $\Omega$& $2.60$         & $1.99$              & ... \\
Albedo                                      & 1.4 $\pm$ 0.14  & 0.62 $\pm$ 0.05 & ...\\
Limb darkening law                &  Square root$^{\star}$ & Square root$^{\star}$ & ...\\
N1, N2                                      & 90$^{\star}$  & 90$^{\star}$ & ...\\
N1L, N2L                                    & 60$^{\star}$  & 60$^{\star}$ & ...\\

\hline
\multicolumn{4}{l}{\textbf{Notes:} $^{\star}$ Fixed, $^1$ \cite{Lehmann2013}}\\
\end{tabular}
\end{table*}

\section{Frequency analysis}
\label{sec:freqanalysis}
Apart from the eclipses, the light curve of KIC\,10661783 shows clear additional
variability (see Fig.\,\ref{lc}) that mainly can be attributed to pulsations. In order to extract frequencies, we subtracted our eclipsing model from the original data and analysed the residua. We calculated amplitude spectra by means of a discrete Fourier transform \citep{Deeming1975,Kurtz1985} and followed the standard pre-whitening procedure.

Given the total number of points in the analysed SC light curve, performing the Fourier analysis up to the Nyquist frequency ($\sim 730$\cpd) is very time consuming.
Since our preliminary study of the periodograms calculated for SC data light curve showed numerous frequency peaks, and none over the $200$\cpd, therefore we decided to stop calculating periodograms on the original light curve at $200$\cpd for both, SC and LC data.

\begin{figure}
    \centering
    \includegraphics[width=\columnwidth,clip]{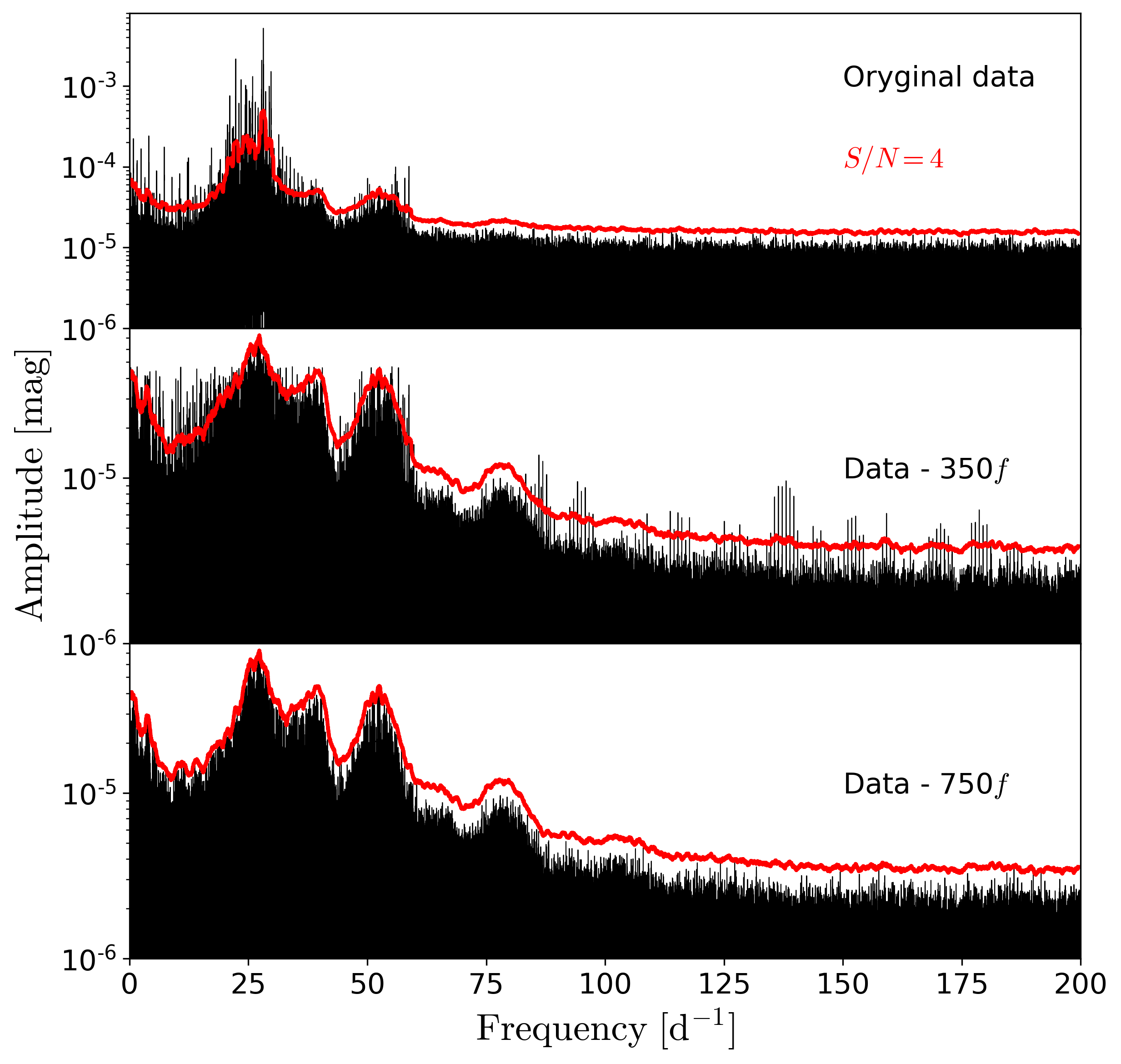}

    \caption{Periodograms calculated for the \textit{Kepler} SC observations corrected for the binary orbit.
    The amplitude spectra for the original data are shown in the top panel.
    The middle and bottom panels present the periodograms calculated for the light curve pre-whitened for 350 and 750 frequencies, respectively. The 4\,S/N level is marked with the red lines. Note that the Y-axis scale differs between the panels.}
    \label{periodograms}
\end{figure}

We assumed the signal-to-noise ratio $\rm S/N=4$ as a threshold for significant frequencies \cite[see][]{Breger1993,Kuschnig1997}, however those with S/N<5 should be treated with caution \citep{Baran2015}.
The noise was calculated as an average amplitude value in a 1\cpd\ window centred at a given peak before its extraction.

Our careful analysis has revealed 750 frequency peaks for the SC data. Fig.\,\ref{periodograms} presents the periodograms calculated for the original SC data corrected for the binary orbit (the top panel), after pre-whitening for 350 (the middle panel) and for 750 found frequencies (the bottom panel). The residual periodogram, presented on the bottom of Fig.\,\ref{periodograms}, has some visible humps around 25, 35, 50 and 75\,\cpd.
Those humps are most probably due to the unresolved signal left in the data.
Many of the frequencies detected in the SC light curve are present in the LC data set as well.
Naturally, in the case of the LC observations there is a problem with aliasing due to low value of the pseudo-Nyquist frequency ($\sim 25$\cpd), however the LC data gives better frequency resolution and extracted frequencies can be compared with those found from the SC analysis.

In the next step, we applied a selection  criterion to these 750 frequencies using the resolution condition. We adopted the resolution of 1.5 times the Rayleigh limit i.e., 1.5/T, where T is the total time span of the observations \citep{Loumos1978}. This gives the values of $\Delta f_{\rm R, SC} = 0.00195$\cpd\ and $\Delta f_{\rm R, LC} = 0.00102$\cpd\ for SC and LC data, respectively.
Then, we checked whether frequencies found in the SC are separated by the distance lower than $\Delta f_{\rm R, SC}$. If so, and both frequencies have their equivalents in the LC data (with the accuracy of $\Delta f_{\rm R, SC}$) with the separation greater than $\Delta f_{\rm R, LC}$ then we accept those frequencies as real ones. If their separation is less than $\Delta f_{\rm R, LC}$, we treated the frequency with lower amplitude as spurious and remove it from the list.
After this procedure, we rejected 160 frequencies from the SC. The remaining set of 590 significant frequencies we regard as a final one for the further identification of possible combinations.

\begin{figure}
    \centering
    \includegraphics[width=\columnwidth,clip]{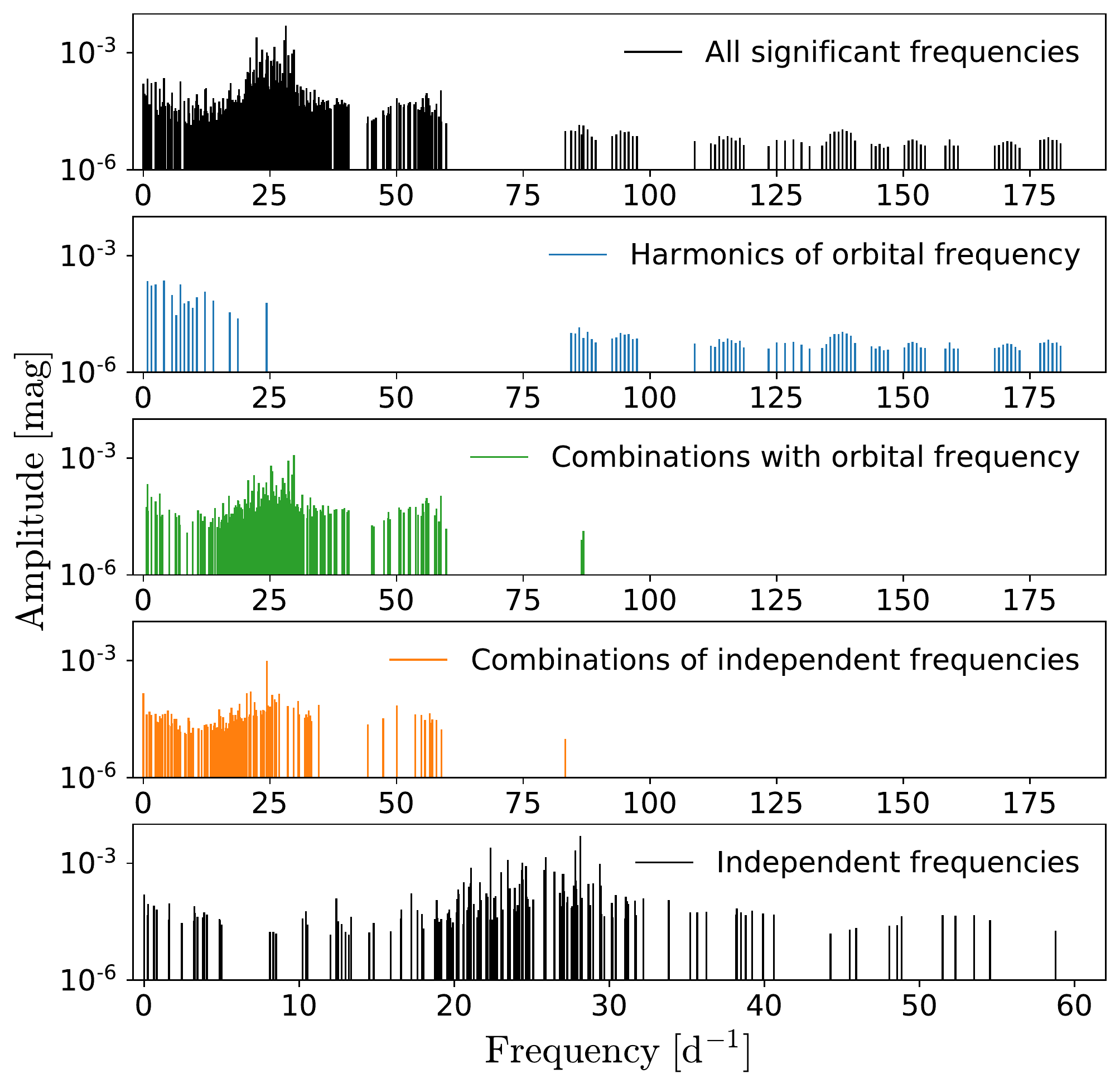}

    \caption{The frequency peaks from the analysis of the SC data after subtraction of the orbital model and after a selection for the frequency resolution. In the top panel, we show all 590 significant frequencies found in the SC data.
    The lower three panels show:  the orbital frequency harmonics, combinations with the orbital frequency and combinations between the independent frequencies.
    The bottom panel shows only the independent frequencies (207 peaks). Note that the X-axis scales differ between the panels.}
    \label{osc}
\end{figure}

We looked for possible combinations of all significant frequencies and the orbital frequency using simple formula: $m \times f_i + n \times f_j$, with $m$ and $n$ being integers between -10 and 10. Moreover, we identified the orbital harmonics, i.e.,  $N \times f_{\rm orb}$, with N being an integer
greater than zero. Finally, we determined that 207 amongst all significant frequencies seem to be independent within the adopted frequency resolution.
In Fig.\,\ref{osc}, we show the final results of the frequency analysis. In the five panels from the top to bottom, we can see: all significant frequency peaks found in the SC data, harmonics of the orbital frequency (83 peaks), combinations with the orbital frequency, combinations of independent frequencies and, in the bottom panel, only the independent frequency peaks.
A complete list of the frequencies after the rejection of those regarded as spurious due to the adopted frequency resolution
is in Appendix\,\ref{sec:appendix}. The possible combinations are listed in the Remarks column.

\begin{figure}
    \centering
    \includegraphics[width=\columnwidth,clip]{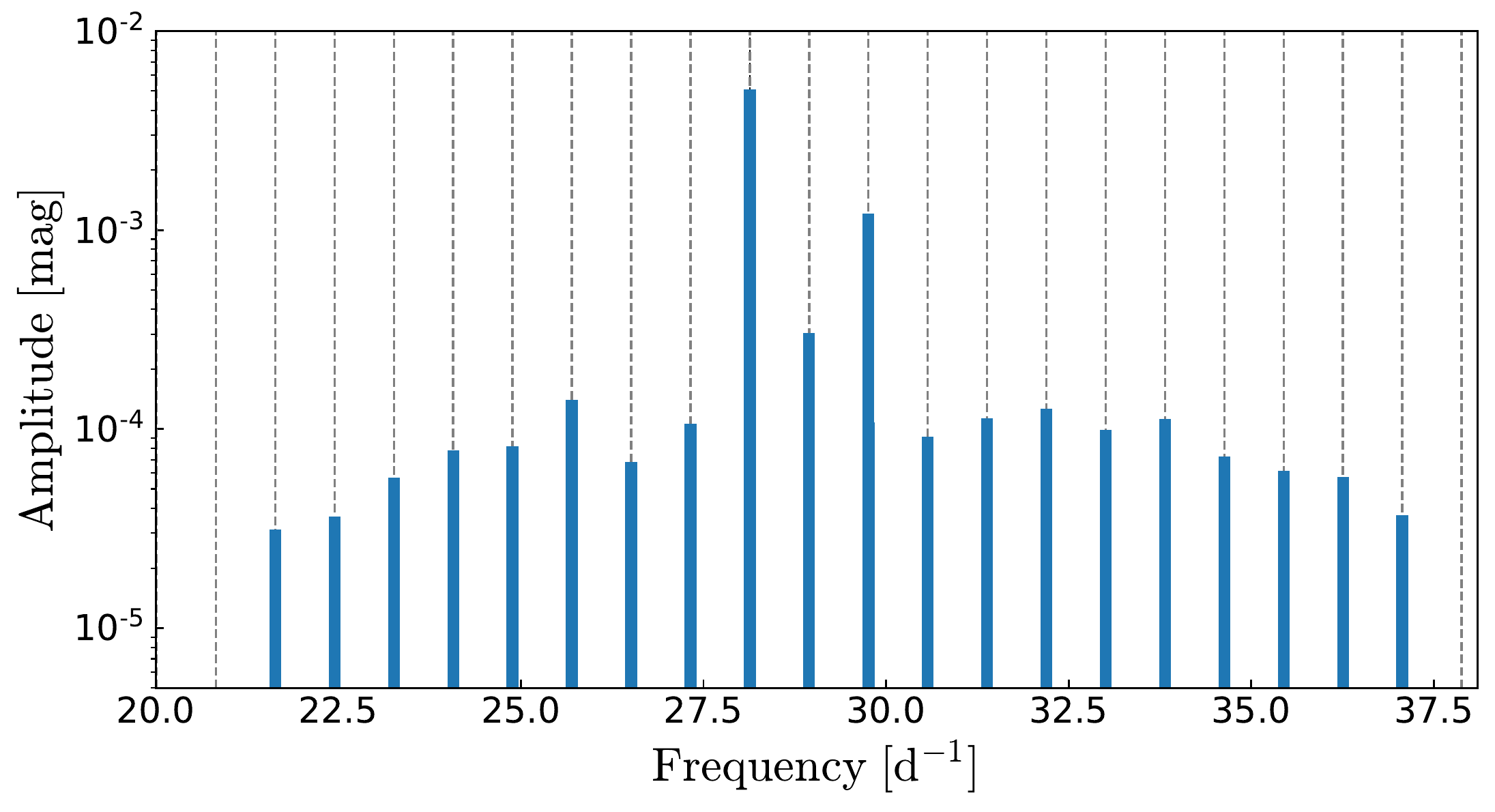}

    \caption{Combinations of the dominant pulsational frequency with the orbital frequency ($f_1 + N\times f_{\rm orb}$). All frequency peaks are equidistant by the orbital frequency with the accuracy of $\Delta f_{\rm R, SC}$.}
    \label{f1_combinations}
\end{figure}

Despite the fact, that the selected frequencies range up to $\sim$ 180\cpd, above $80$\cpd\ we observe only harmonics of the orbital frequency (up to $223 \times f_{\rm orb}$). Such frequencies can appear whenever one considers a light curve corrected for imperfect binary model or when a tidally-locked pulsations occur.
What is more, we also found combinations of the pulsation frequencies and the orbital frequency which may appear when
the amplitude depends on which side of the star is oriented towards the observer (i.e. on the orbital phase). The second plausible explanation is that, since we analyse the system undergoing eclipses, the component's contribution to the total light changes with the orbital period. Even if the primary component would pulsate with the constant amplitude, the observed amplitude would change during eclipses. In Fig.\,\ref{f1_combinations}, we show the combinations of the strongest pulsational frequency with the orbital frequency. 

\begin{figure}
    \centering
    \includegraphics[width=\columnwidth,clip]{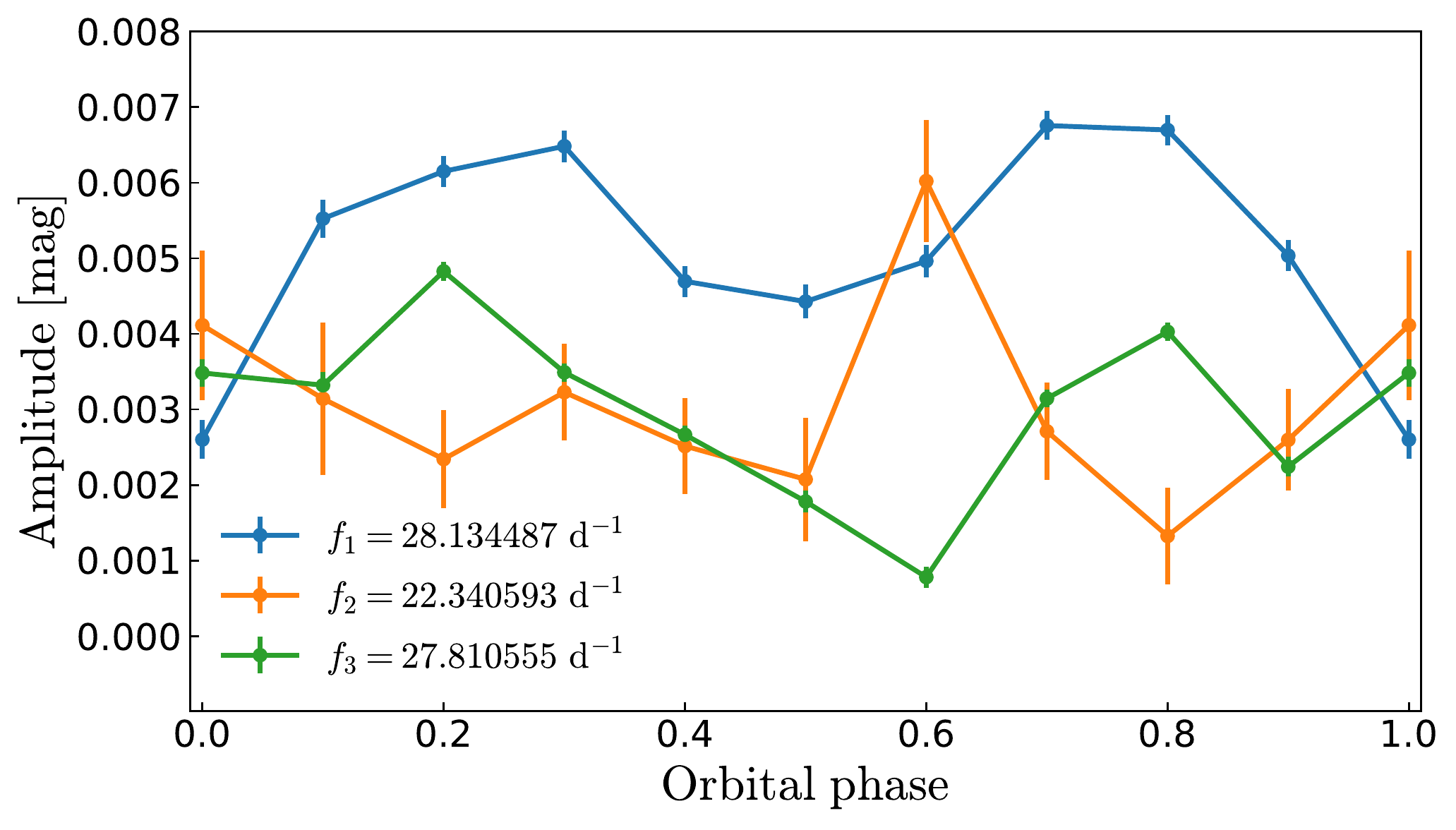}

    \caption{The amplitude modulation effect presented for three strongest, independent frequencies. Details are described in the text.}
    \label{ampl_modulation}
\end{figure}

To investigate this phenomenon closely, we removed all periodicities from the light curve except for $f_1$ and its combinations with the orbital frequency.
Next, we divided the light curve into intervals in the orbital phase $\Delta \phi = 0.1$ and fitted the frequency $f_1$ to determine the amplitudes and phases
in each interval separately. Such procedure was also repeated for $f_2$ and $f_3$.
The amplitude change can be seen in Fig.\,\ref{ampl_modulation}, where we plotted the amplitudes for three strongest modes as a function of the orbital phase.
As one can see, these amplitudes change with the orbital phase. In the case of $f_1$, it resembles sinusoidal variability with the amplitude maxima occurring near
the orbital phase 0.3 and 0.7.
Moreover, the amplitude of $f_2$ seems to be in anti-phase with the amplitudes of $f_1$ and $f_3$.

Such amplitude modulation can suggest that one may be able to distinguish various parts of the primary with different values of the pulsational amplitude. Such a case was explored by \cite{Springer2013} where the authors studied pulsational distribution on the tidally deformed component of the binary system. They concluded that the more the star is tidally deformed the more pulsations are trapped in the circumpolar region located in the hemisphere that is facing outwards the system. Their theoretical results resemble our observations. Only recently \cite{Handler2020} reported the first ever found binary system with a star that pulsates only on the one hemisphere, facing either the first or third Lagrange point ($L1$ or $L3$).
CO Cam is the second reported case of a system exhibiting pulsations on a hemisphere facing the $L1$ point at least in four frequencies \citep{Kurtz2020}.
\cite{Fuller2020} explained this phenomenon  as a result of tidal mode coupling and called it \textit{tidally tilted pulsators}.
KIC\,10661783 may be another star pulsating in a similar way.

\section{Binary evolution models}
\label{sec:binaryevolution}

Binary systems present a wide variety of interactions altering the evolution of both components and the system as a whole. Mass-transfer events causing the rejuvenation of one of the components is only one of many effects that are crucial to be taken into account when modelling the stellar evolution.

To model the binary KIC\,10661783 as a product of a binary evolution, we used the Modules for Experiments in Stellar Astrophysics \citep[\texttt{MESA},][]{Paxton2011, Paxton2013, Paxton2015, Paxton2018, Paxton2019}
in its 12115 version  with the \texttt{MESA-binary} module.  \texttt{MESA} relies on the variety of the input microphysics data.
The \texttt{MESA} EOS is a blend of the OPAL \citep{Rogers2002}, SCVH \citep{Saumon1995}, FreeEOS \citep{Irwin2004}, HELM \citep{Timmes2000}, and PC \citep{Potekhin2010} EOSes.
Radiative opacities are primarily from the OPAL project \citep{Iglesias1993,Iglesias1996}, with data for lower temperatures from \citet{Ferguson2005} and data for  high temperatures, dominated by Compton-scattering from \citet{Buchler1976}. Electron conduction opacities are from \citet{Cassisi2007}.
Nuclear reaction rates are from JINA REACLIB \citep{Cyburt2010} plus additional tabulated weak reaction rates from \citet{Fuller1985}, \cite{Oda1994} and \cite{Langanke2000}. Screening is included via the prescription of \citet{Chugunov2007}. Thermal neutrino loss rates are from \citet{Itoh1996}.
The \texttt{MESA-binary} module allows to construct a binary model and to run and compute the evolution of both components including such effects as mass transfer and orbital elements evolution.
Roche lobe radii in binary systems are computed using the fit of \citet{Eggleton1983}. Mass transfer rates in Roche lobe overflowing binary systems are determined following the prescription of \citet{Ritter1988}.

For our evolutionary compuations, we used \cite{Asplund2009} chemical mixture and OPAL opacity tables. We adopted the Ledoux criterion for the convective instability with the mixing-length theory description by \cite{Henyey1965} and semi-convective mixing with the parameter $\alpha_{\rm SC}=0.01$. The diffusive exponential overshooting scheme described by \cite{Herwig2000} and parametrized by the parameter $f_{\rm ov}$ was applied.
For the large-scale effects we used the mass transfer of Kolb type \citep{Kolb1990}. We included stellar winds from both components following the prescription of \cite{Vink2001}. For the sake of simplicity of computations we ignored the rotation of stars.

Whenever one considers binary evolution the crucial point is a choice between conservative and non-conservative mass transfer scheme, which means that a mass is not lost or lost from the system during the transfer, respectively. However, it is still discussed which type should be considered \cite[see e.g.,][]{Kolb1990,Sarna1992,Sarna1993,Guo2017}. Recently, \cite{Chen2017} concluded that the formation of stars similar to KIC\,10661783 can be fully explained in non-conservative way. Here, we rely on the non-conservative transfer described as the fraction of mass lost from the vicinity of the accretor in the form of a fast wind during the mass transfer \citep{Tauris2006}. This mass fraction is denoted by the parameter $\beta$, with the value varying between 0, meaning totally conservative mass transfer, and 1 describing totally non-conservative mass transfer scheme.

\begin{figure*}
    \centering
    \includegraphics[width=1.75\columnwidth,clip]{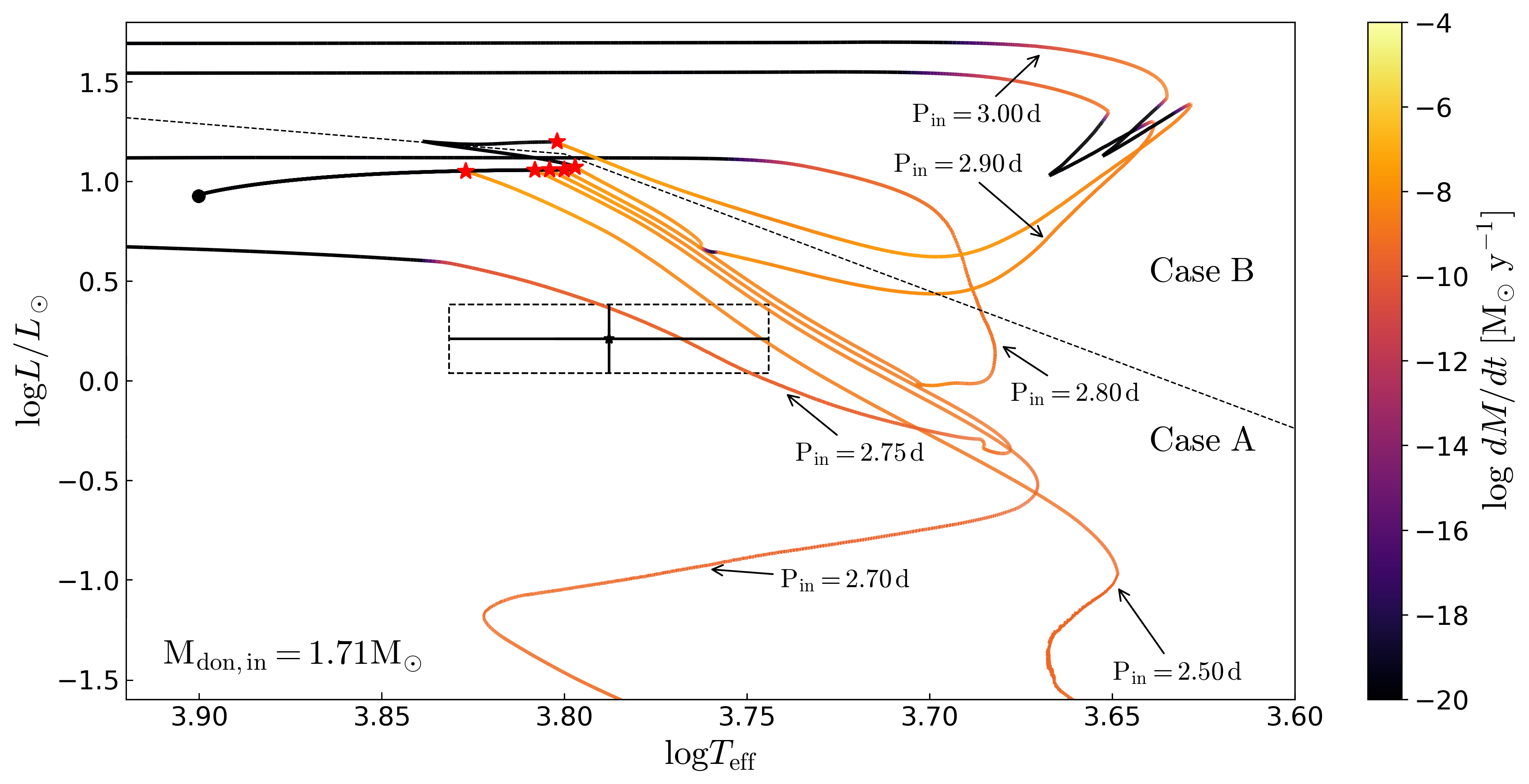}
    \caption{The Hertzsprung-Russel diagram with the donor evolutionary tracks computed as a results of the binary evolution of components with the initial masses $M_{\rm don,ini}=1.71$\,\mass\ and $M_{\rm acc,ini}=1.15$\,\mass and with six values of the initial orbital period, $P_{\rm in}$. The other input parameters are: $X_0=0.70$, $Z=0.020$, $f_{\rm ov}=0.01$, $\alpha_{\rm MLT}=2.0$ and $\beta=0.5$. All tracks start from a black dot. The beginning of the mass transfer is marked with the red stars. The colour of the tracks depends on the mass transfer rate at a given evolution moment, as specified in the colour-bar. The dashed line separates the two mass transfer scenarios: case A and B. The position of the secondary component of KIC\,10661783 is marked with the 3$\sigma$ error box.}
    \label{P_dependence}
\end{figure*}

Except of the initial masses, the main parameter controlling the evolution of the binary undergoing the mass-transfer is the initial orbital period, $P_{\rm in}$. It is also a main parameter that distinguishes between \textit{A} and \textit{B} Roche-lobe overflow cases \citep{Kippenhahn1967}. Case \textit{A} describes the mass exchange with a donor in the main-sequence phase and case \textit{B} the mass exchange in the rapid core contraction phase preceding helium ignition. We followed the evolution of both components with the initial orbital periods between 1.8 to 4.0\,d with the step reaching the value down to $\Delta {P}=10^{-5}$\,d. In order to demonstrate the importance of the orbital period on the systems evolution, in Fig.\,\ref{P_dependence} we show evolutionary tracks of the 1.71\mass\ donor, that evolves in a binary system with the 1.15\mass accretor. Calculations were done for $Z=0.020$ and $X_0=0.70$. All tracks start from a black dot. It is clearly visible that the moment when the mass transfer starts (marked with red stars) is fully controlled by the initial orbital period.

In order to obtain a model reproducing the system at its current evolutionary stage, with masses and radii as determined by \cite{Lehmann2013} and the orbital period from our analysis, we built an extensive grid of models\footnote{For inlists see \url{https://doi.org/10.5281/zenodo.4618112}}. Our grid covered a wide range of initial masses; 0.7 to 1.7\mass\ for the accretor and 1.0 to 2.5\mass\ for the donor, with the step descending iteratively, down to the value of $\Delta \rm{M}=0.001$\mass, whenever a local minimum of fitting was found.

Firstly, a sparse grid of models for $Z=0.014$ and $X_0=0.7$ was computed for aforementioned mass and period ranges.
We tested different values of $\beta$ (between 0 and 1 with $\Delta \beta$=0.1) and $f_{\rm ov}$ (between 0.00 and 0.04, with $\Delta f_{\rm ov}$=0.01) to find their preferable values.
To find the best models, for each star we calculated the discriminant $D^2$:
$$ D^2 = \frac{1}{N}\sum_{\rm i=1}^{\rm N} \left( \frac{X_{\rm obs, i}-X_{\rm model, i}}{\sigma_{\rm obs, i}} \right)^2$$
where $X$ denotes the considered parameters, i.e.,  the orbital period, mass, radius, effective temperature and luminosity for each star. $N$ is the total number of considered parameters.
Then, the mean value of $D^2$ was computed as
$$<D^2> = \frac{D_1^2 + D_2^2}{2},$$
where $D_1$ and $D_2$ are the discriminants for the primary and secondary, respectively.

In the first step, from the computed grid, we selected the models, that for a given value of $P_{\rm in}$ reproduce masses and radii of both components within 3$\sigma$. The most preferable initial value of the accretor mass was $0.9$\mass\ while the initial donor mass was about 1.4\mass. The totally conservative mass transfer ($\beta=0.0$) and the overshooting $f_{\rm ov}=0.02$ were preferred. However as we noted, our models had difficulties with reproducing the observed radius of the donor.
In order to fix that discrepancy, we tested the dependence between the final masses and radii of both components on additional parameters, like metallicity $Z$, initial hydrogen abundance $X_0$ and mixing length parameter $\alpha_{\rm MLT}$. We found that increasing metallicity to $Z=0.025$, the initial hydrogen abundance to $X_0=0.73$
and adopting $\alpha_{\rm MLT}=1.3$ helps to improve the agreement. However for such parameters  we had to move away from the conservative mass transfer assumption
 to $\beta=0.05$, i.e. 5\% of the transferred mass being lost from the system.

\begin{figure}
    \centering
    \includegraphics[width=\columnwidth,clip]{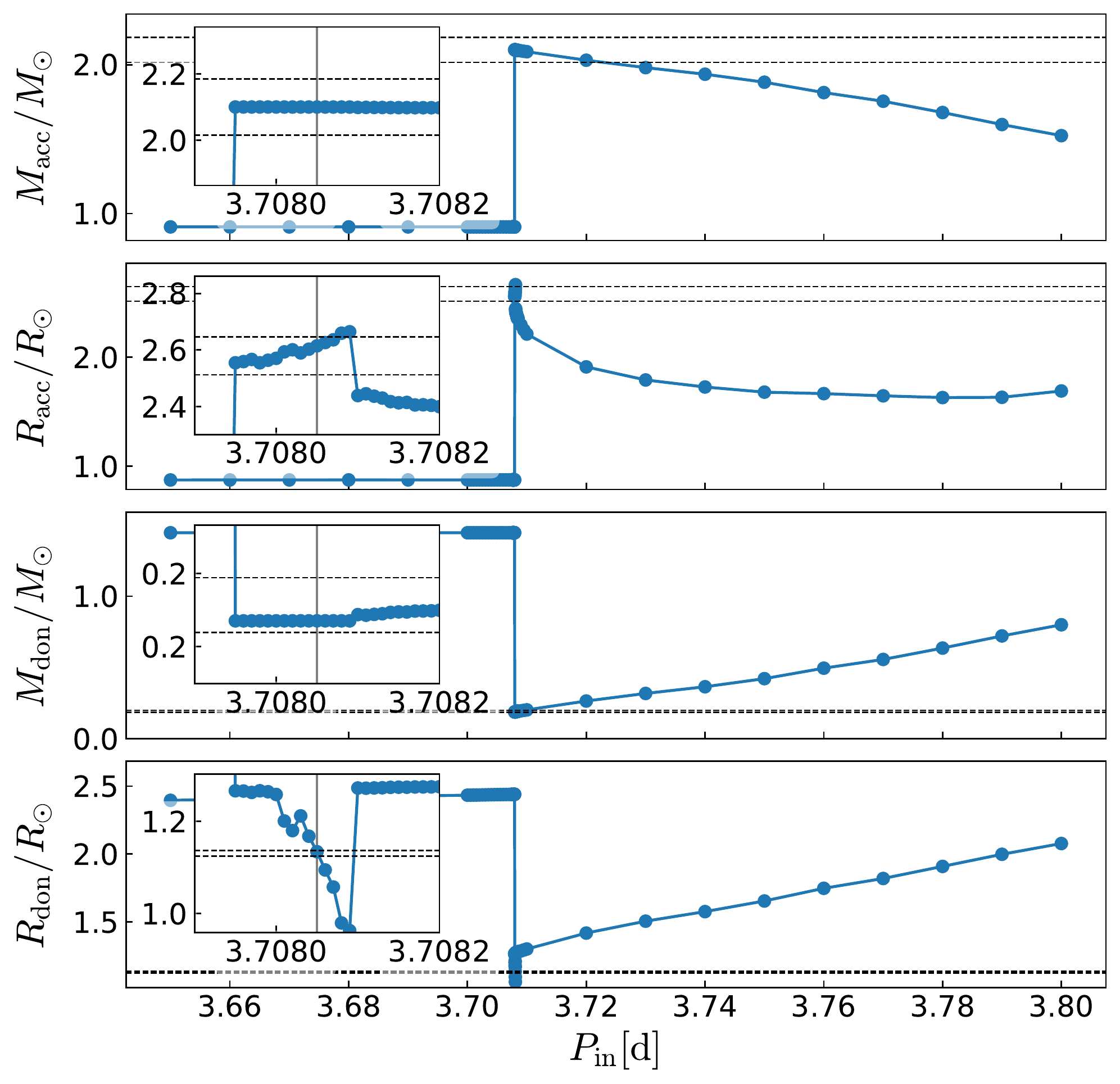}
    \caption{The dependency of the final masses and radii of each component on the initial orbital period $P_{\rm in}$. The models have parameters: $M_{\rm don,in}=1.45$\mass, $M_{\rm acc,in}=0.91$\mass, $Z=0.025$, $X_0=0.73$, $f_{\rm ov}=0.02$, $\alpha_{\rm MLT}=1.3$ and $\beta=0.05$. The horizontal lines give the observed $3\sigma$ range of masses and radii. The vertical, grey lines on the insets mark the best solution.}
    \label{R_M}
\end{figure}

In order to find the best model we used the already found preferable values of the parameters and changed the approach. Instead of making a finer mesh of models, we used diagrams showing the dependence of the final values of masses and radii on the initial period, $P_{\rm in}$. The examples of such diagrams are presented in Fig.\,\ref{R_M}, where we plotted models varying only in the initial period $P_{\rm in}$, while the following parameters have been fixed: $M_{\rm don,ini}=1.45$\mass, $M_{\rm acc,ini}=0.91$\mass, $Z=0.025$, $X_0=0.73$, $f_{\rm ov}=0.02$, $\alpha_{\rm MLT}=1.3$ and $\beta=0.05$. As one can see, there is a clear dependence of the final masses and radii on the initial period. Too high value of $P_{\rm in}$ allows the donor to reach the red giant phase, after which it enters the nearly-constant luminosity phase leading further to several flashes in a H-burning shell before it cools down. In such case, the orbital period begins to increase rapidly, at the onset of mass transfer, reaching values of tenths of days in a cooling phase.
On the other hand, too low value of $P_{\rm in}$ causes a slow rate of mass transfer, at which the donor's radius and the systems orbital period are decreasing. However, the donor is not able to contract to become the helium white dwarf (He-WD). The boundary between those scenarios is set by the so-called \textit{bifurcation period}, at which there is a sudden drop, visible in Fig.\,\ref{R_M}.

\begin{figure*}
    \centering
    \includegraphics[width=1.75\columnwidth,clip]{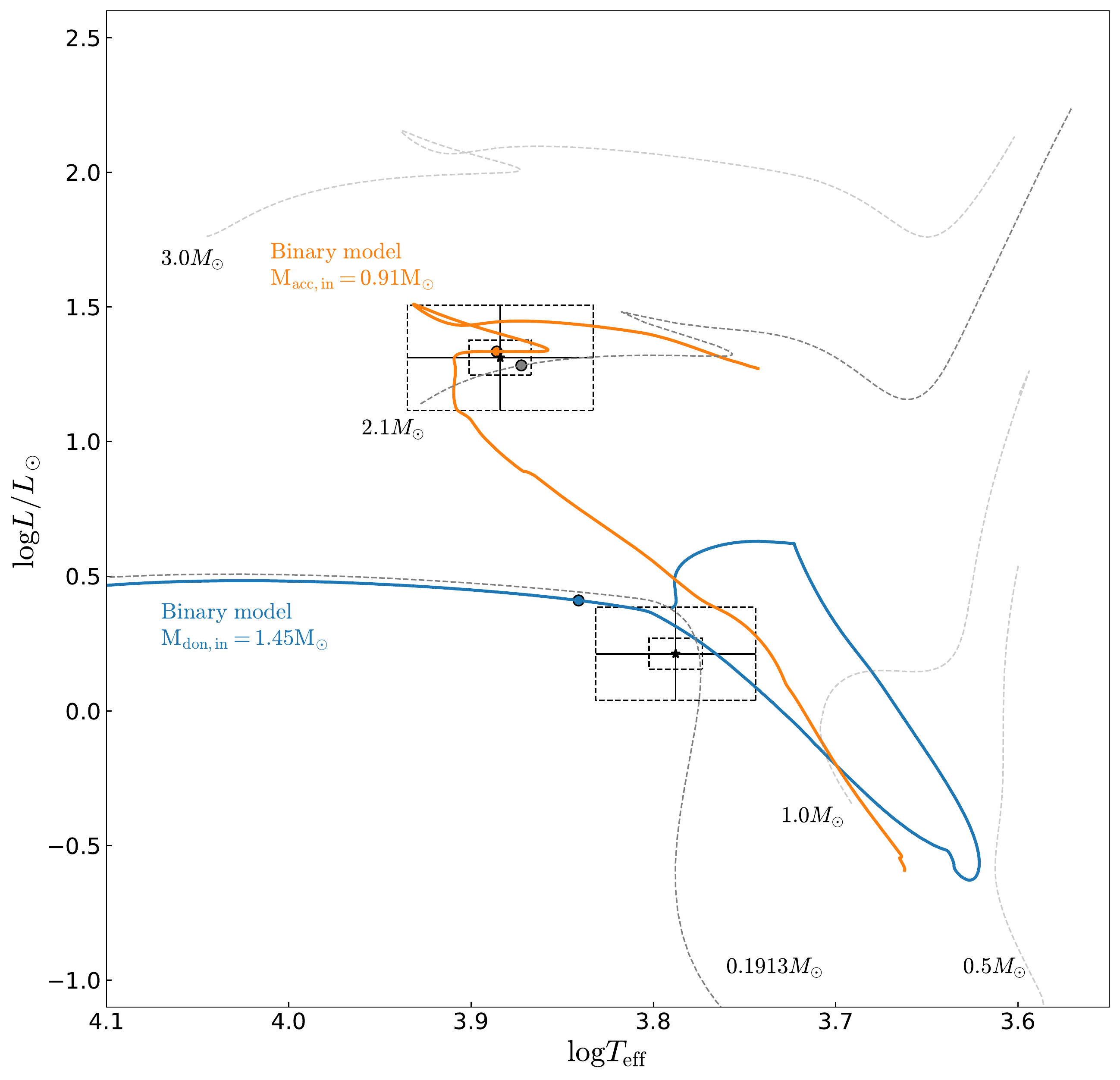}
    \caption{Binary evolutionary tracks in the HR diagram computed for the best matching set of parameters, as described in Table\,\ref{tab:parameters}. The orange and blue tracks mark binary evolution models for the accretor and donor, respectively. Grey evolutionary tracks calculated from single evolution of various initial masses are plotted for a comparison. Thick grey lines represent tracks for masses corresponding to the current masses of the binary components. Positions of the primary and secondary components of KIC\,10661783  are marked with 1 and 3$\sigma$ boxes. Orange dot, inside the 1$\sigma$ box of the primary component, marks the position of the best model reproducing the mass, radius, effective temperature and luminosity. Grey dot represents the position of the model from single evolution. Blue dot marks the position of the donor model from binary evolution.}
    \label{HR_best_models}
\end{figure*}

The parameters of the best model (in terms of the discriminant) reproducing the orbital period, masses and radii of both components as well as their positions in the HR diagram are presented in Table\,\ref{tab:parameters}. However, the secondary component in our best model lies slightly outside 3$\sigma$ error box in the HR diagram.
The model has $M_{\rm don,ini}=1.45$\mass, $M_{\rm acc,ini}=0.91$\mass and the initial period, $P_{\rm in}=3.70805$\,d.
In Fig.\,\ref{HR_best_models}, we plotted the corresponding evolutionary tracks for the accretor (the orange line) and donor (the blue line) in the HR diagram. For a comparison, we show also evolutionary tracks calculated for a single star evolution with grey lines, adopting the same values of $Z$, $X_0$, $f_{\rm ov}$ and $\alpha_{\rm MLT}$ as in the binary case. The positions of the primary and secondary components are shown with their 1$\sigma$ and 3$\sigma$ errors, inside which we marked the best fitting models.

We found that in order to fit the secondary component within its observed ranges of mass and radius, we must ensure that the donor looses most of its outer shell before it evolves towards the red giant phase. For such mass sets, describing both initial and current state of KIC\,10661783 system, this can be obtained only in the case A, i.e., with mass transfer occurring in the main sequence phase of the donor. In this scenario, during the mass transfer, donor enters the nearly constant-luminosity phase with ever-growing effective temperature, towards H-shell flashes, to the cooling sequence of the helium dwarf stage.

\begin{figure}
    \centering
    \includegraphics[width=\columnwidth,clip]{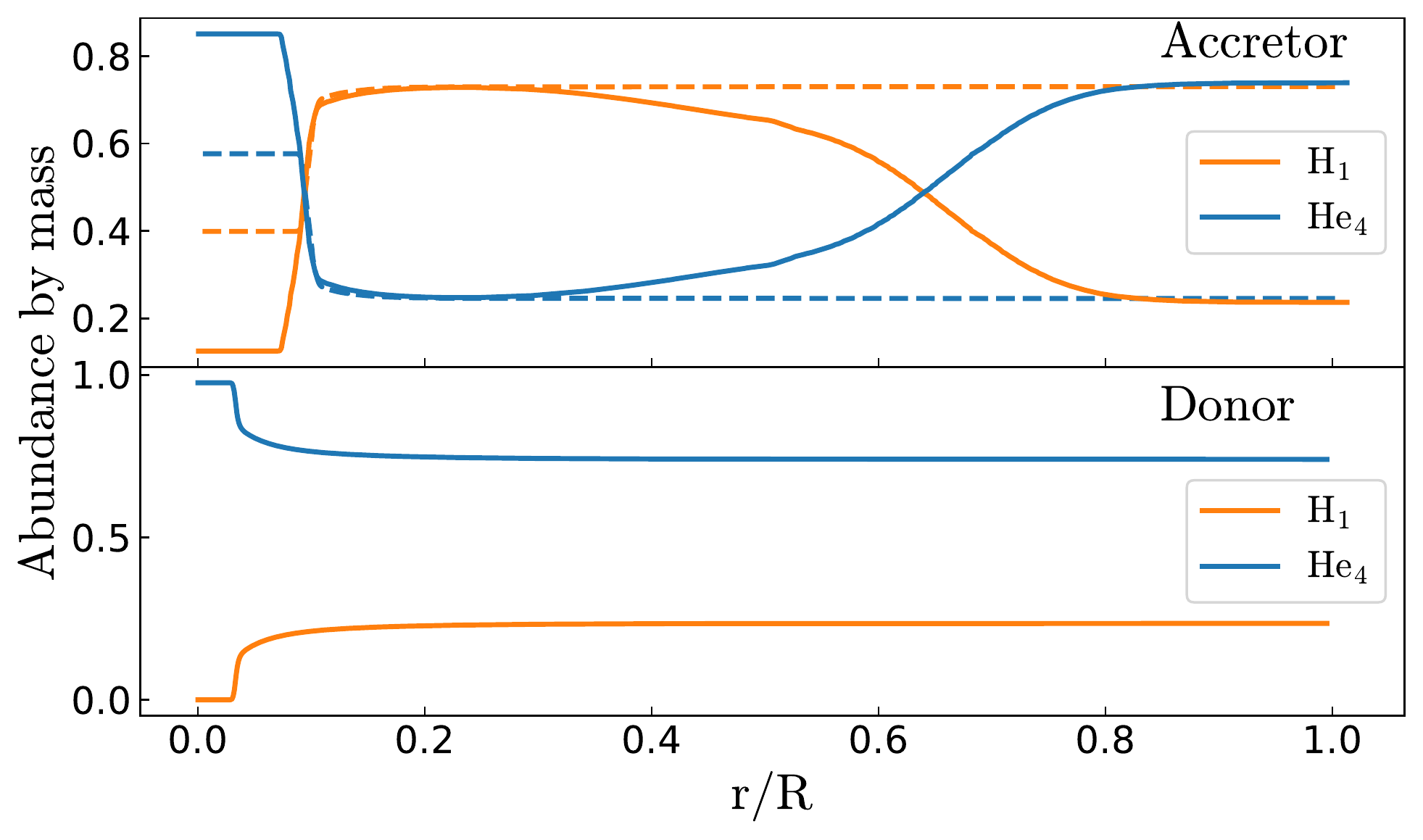} \\
    \includegraphics[width=\columnwidth,clip]{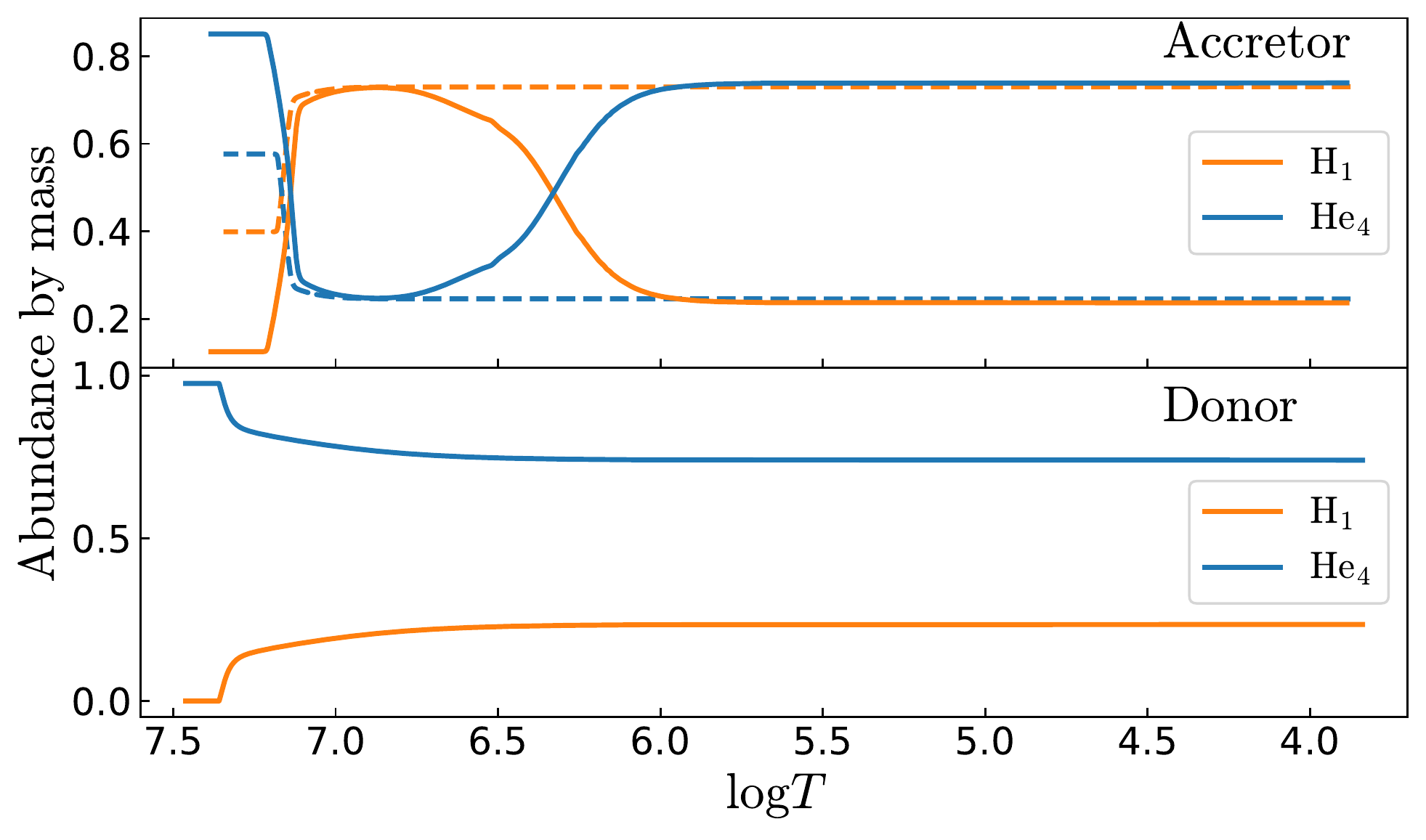}
    \caption{The abundance profiles of H and He as a function of the relative radius (the top panel) and temperature (the bottom panel) for both components. The profiles obtained from the the binary-evolution computations are plotted with solid lines whereas the profiles resulting from the single evolution with dashed lines. The primary mass for, both, single and binary-evolution computations is 2.1\mass, while the secondary has a mass of 0.187\mass.}
    \label{abundance_profiles}
\end{figure}

In the next step, we studied the effect of binary evolution on the internal abundance profiles of H and He of both components. In Fig.\,\ref{abundance_profiles}, we present the H and He profiles as a function the relative radius (the top panels) and temperature (the bottom panels) for both components for the best binary model we found. For a comparison, we show the abundance profiles for the single-evolution model of the primary with dashed lines. This model was calculated for the observed value of the primary mass, i.e., $M=2.1$\mass, using the same set of parameters,
i.e. $Z=0.025$, $X_0=0.73$, $\alpha_{\rm MLT}=1.3$, $f_{\rm ov}=0.02$, and assuming the same prescription and parametrisation of stellar wind.
This model is marked in Fig.\,\ref{HR_best_models} with the grey dot.
Since the single primary's model undergoes a standard main sequence evolution, it has an extended hydrogen envelope surrounding a H-burning core enriched in helium up to 60\%. Its binary equivalent manifests the influence of the past-mass transfer history on its interior. Because it is much older than the corresponding single model (6.39 vs 0.97 Gyr), it has much higher abundance of He in the core. The outer shell reveals interesting property of the H and He profiles. Near the core boundary its structure reflects the initial hydrogen and helium composition, however the further away from the core, the higher abundance of helium relative to the hydrogen abundance. At the surface
we got the reversed ratio of H/He. It results in the outermost He rich layer being a direct manifestation of mass transfer in the past.
The secondary, on the onset of mass transfer, is still fusing H to He. By that time, which is nearly 4.7 Gyr, it was able to create almost fully-He core (of about 92\%) with only 5\% of H left. That means, that the mass-transfer event happened right before the overall contraction. Although it is systematically being stripped out of its outer layers, the inner-core fusion is still producing He. The moment of the core conversion  to fully helium and the beginning of the shell H-burning corresponds roughly to the minimum of effective temperature and luminosity of the donor track in Fig.\,\ref{HR_best_models}. Since now, the luminosity and effective temperature of the donor increase, due to the new region of H-burning, leading it to the current observed status as a helium-core pre-white dwarf.
The profile of the metallicity $Z$ is almost constant throughout the interior of both components and almost independent of the evolutionary past of the system.

As determined by \citet{Lehmann2013} from spectroscopy, KIC\,10661783 exhibits some anomalies in the abundances of elements like C, N and O, what can be explained
by the mass transfer in the past. From their determinations, the abundance of nitrogen is much greater than the solar value of \citet{Asplund2009} (AGSS09),
the abundance of oxygen is about the same, and carbon is less abundant than the value of AGSS09.
The abundances of N and O from our binary-evolution modelling agree with the results of \citet{Lehmann2013} within the 2$\sigma$ error, while the abundance of C of the primary
is below $4\sigma$ of the Lehmann's value.
Our abundances of CNO are given in Table\,\ref{tab:abundances} together with determinations of \citet{Lehmann2013} and the AGSS09 solar values for a comparison.

\begin{table*}
    \caption{The parameters of the system and both components from our WD and from the analysis of \citet{Lehmann2013}.
    In the last column we give the set of parameters obtained from the best fitting of the evolutionary \texttt{MESA-binary} models.}
    \label{tab:parameters}
    \begin{tabular}{lccc}

        \hline
        \hline
        Parameters      & \texttt{WD}       & \cite{Lehmann2013}            & \texttt{MESA} \\
                               &    Model &  &  Model \\
        \hline

        \multicolumn{4}{l}{\textbf{----- Initial parameters -----}} \\
        Initial orbital period $P_{\rm in}$ (d)    &   --  &   --  & 3.70805  \\
        Donor initial mass (\mass)       &  --  &   --  & 1.450  \\
        Accretor initial mass (\mass)       &   --  &   --  & 0.910 \\
        Initial Z       &   --  &   --  & 0.025 \\
        Initial $X_0$       &   --  &   --  & 0.73 \\

        \multicolumn{4}{l}{\textbf{----- Orbital parameters -----}} \\
        Orbital period P (d)    & \textbf{$1.23136326 \pm 3 \times 10^{-8}$} & \textbf{$1.23136220 \pm 2.4 \times 10^{-7}$} & 1.23136326 \\
        $M_2/M_1$   &   0.09109$^{\star}$   & 0.09109 & 0.08906   \\
        Age (Gyr)   &   --  &   --  & 6.39 \\

        \multicolumn{4}{l}{\textbf{----- Primary star (accretor) -----}} \\
        Mass (\mass)            &   --                & 2.100 $\pm$ 0.028 & 2.100   \\
        Radius (R$_{\odot}$)    & 2.5793 $\pm$ 0.0224 & 2.575 $\pm$ 0.015 & 2.618 \\
        $\log T_{\rm eff}$ (K)  & 3.8840 $\pm$ 0.0170 & 3.890 $\pm$ 0.003 & 3.886 \\
        $\log L/L_{\odot}$      & 1.3113 $\pm$ 0.0651 & 1.335 $\pm$ 0.013 & 1.335  \\

        \multicolumn{4}{l}{\textbf{----- Secondary star (donor) -----}} \\
        Mass (\mass)            &   --                & 0.1913 $\pm$ 0.0025 & 0.187   \\
        Radius (R$_{\odot}$)    & 1.1320 $\pm$ 0.0020 & 1.124  $\pm$ 0.019  & 1.125  \\
        $\log T_{\rm eff}$ (K)  & 3.7878 $\pm$ 0.0146 & 3.778  $\pm$ 0.007  & 3.838 \\
        $\log L/L_{\odot}$      & 0.2121 $\pm$ 0.0575 & 0.161  $\pm$ 0.026  & 0.407 \\

        \hline
        \multicolumn{4}{l}{\textbf{Notes:} $^{\star}$ Fixed}\\
    \end{tabular}
\end{table*}

\begin{table}
    \centering
    \caption{Element abundances for both primary and secondary component compared to the Sun values, which are given below the element designation.}
    \label{tab:abundances}
    \begin{tabular}{lccc}

        \hline
        \hline
            & C & N & O \\
        Solar \citep{Asplund2009} & 8.43 $\pm$ 0.05 & 7.83 $\pm$ 0.05 & 8.69 $\pm$ 0.05 \\
        \hline

        \multicolumn{4}{c}{\textbf{------- \cite{Lehmann2013} (spectroscopy)---------}}\\
        Primary                 & 8.21 $\pm$ 0.28 & 8.95 $\pm$ 0.34 & 8.6 $\pm$ 0.50 \\
        Secondary               & 7.56 $\pm$ 0.25 & -- & -- \\

        \multicolumn{4}{c}{\textbf{----------- This paper (binary evolution)---------}}\\
        Primary                 & 7.08 & 9.54 & 9.18 \\
        Secondary               & 7.08 & 9.54 & 9.18 \\

        \hline
    \end{tabular}
\end{table}

\section{Pulsation modelling}
\label{sec:PulsationModelling}

The essential step of asteroseismic modelling is mode identification. In the case of one-colour \textit{Kepler} data for KIC\,10661783, with no regularities in frequencies or periods, we were unable to identify any pulsational mode. Therefore, we limited our seismic study to reproducing the mode instabilities in the observed frequency range.

In the modern era of space-based photometric observations, the hybrid stars pulsating in both g- and p-mode regimes are rather a rule than an exception.
In the case of $\delta$ Sct models the low frequencies face a problem because the current standard opacity models predict instability only for higher frequency modes
($f \gtrsim 4$\cpd), while low frequency modes ($f \lesssim 4$\cpd) remain stable \cite[see e.g.][]{Balona2015}.
This can be due to insufficient understanding of the theory of stellar pulsations or due to still existing uncertainties in opacity data.
 The same problem exists in the case of $\beta$\,Cep/SPB hybrid pulsators. This discrepancy cannot be explained by changing model parameters as hydrogen or metal contents in the star as well as by the effects of rotation. To fix that problem, opacity modifications near the Z-bump for $\beta$\,Cep were proposed \citep[e.g.][]{Daszynska2017}.

It was found by \citet{Cugier2012,Cugier2014} in the Kurucz model atmospheres \citep{Castelli2003} that both OPAL and OP opacities are underestimated
near $\log T\approx5.06$\,K.  Guided by this result, \cite{Balona2015} showed that increasing the mean opacity at this temperature
excites low-frequency dipole modes in $\delta$\,Sct models.

\begin{figure*}
    \centering

    \begin{tabular}{cc}
    \multicolumn{2}{c}{\textbf{OPAL opacities}} \\
    \textbf{Single star evolution} & \textbf{Binary evolution} \\
    \includegraphics[width=\columnwidth,clip]{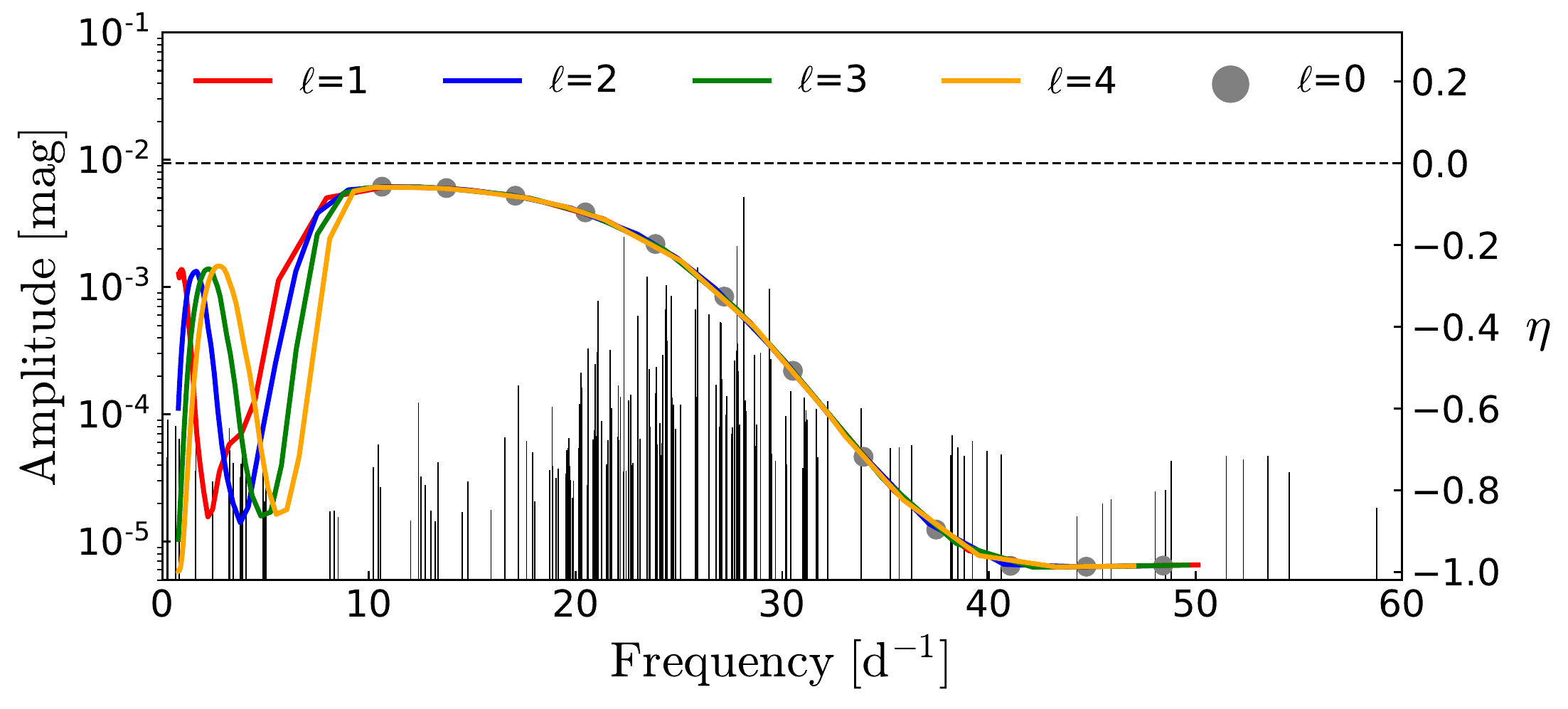}  &  \includegraphics[width=\columnwidth,clip]{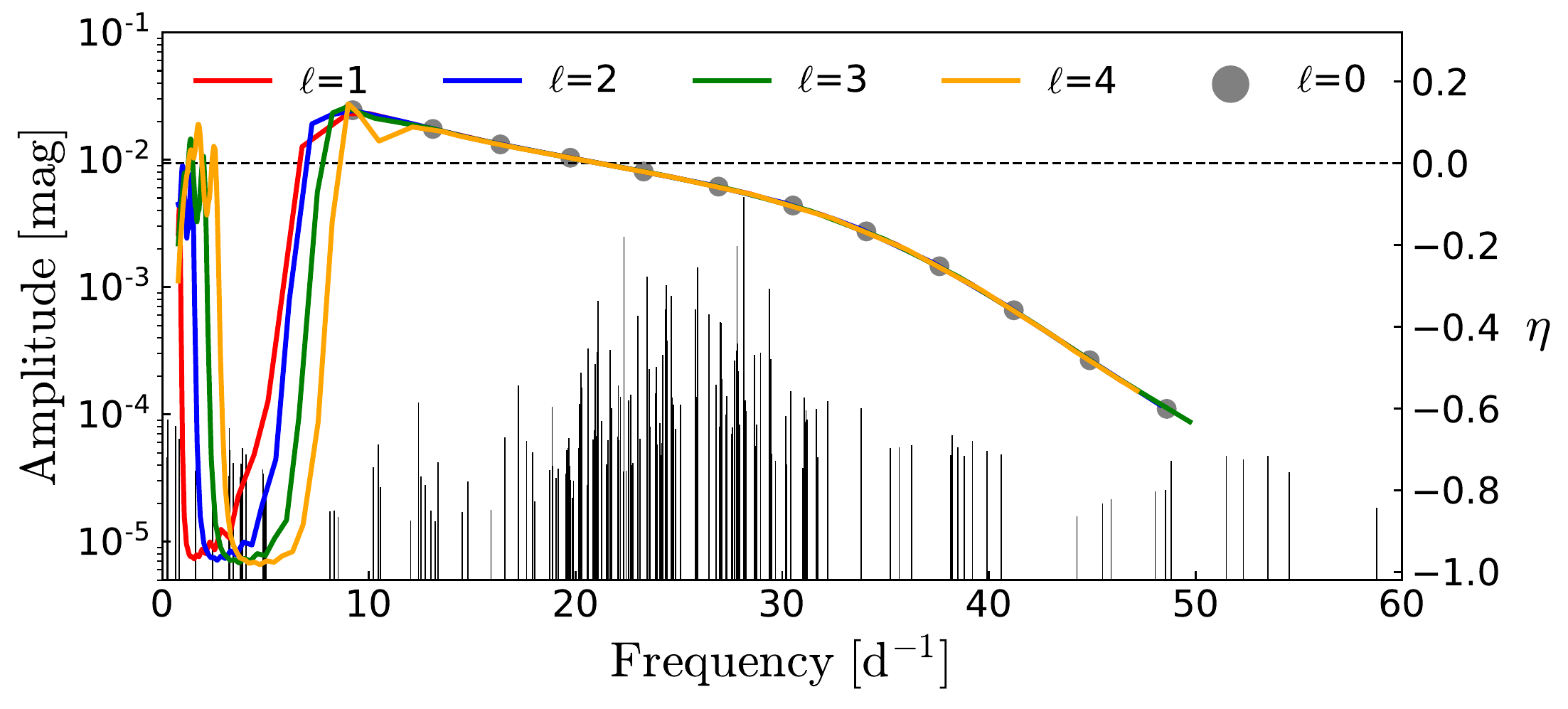}  \\
    \end{tabular}

    \caption{A comparison of the instability parameter $\eta$ between single (the left panel and binary (the right panel) star equilibrium models, for modes with $\ell \le 4$.
    The observed frequency peaks are marked as vertical lines with their amplitudes on the left Y-axis.
    We choose the representative models for each, single and binary case, which fit the measured radius, mass, $T_{\rm eff}$ and $\log L/L_{\sun}$ within 1$\sigma$ errors. Both models
    have very similar parameters (see text) and were calculated for $Z=0.025$, $X_0=0.73$, $f_{\rm ov}=0.02$, $\alpha_{\rm MLT}=1.3$ using OPAL opacity tables. The dashed, horizontal line marks $\eta=0.0$, demarcating the excited ($\eta>0.0$) and suppressed ($\eta<0.0$) modes.}
    \label{opacity_OPAL}
\end{figure*}

For the best model obtained from the binary-evolution computations, we calculated pulsations for the main component using the non-adiabatic code for linear pulsations \citep{Dziembowski1977}. We considered modes with the harmonic degree $\ell = 0 - 4$.
We studied the effect of the binary evolution on the pulsational characteristics by comparing the instability parameter $\eta$ of the binary and single evolution model for the primary component.
The parameter $\eta$, introduced by \cite{Stellingwerf1978}, is a normalised work integral computed over the pulsational cycle. The value of $\eta$ greater than 0 means
that a driving mechanism overcomes damping and the pulsation mode is excited (unstable).
Such comparison can be seen in Fig.\,\ref{opacity_OPAL}, where on the right Y-axis we plotted the instability parameter $\eta$ for representative models of the primary component.
In addition, we also show the observed independent frequencies of the KIC\,10661783 with the values of amplitudes on the left Y-axis.
The single-evolution model, showed in the left panel, has the parameters: $M=2.1$\mass, $R=2.59$\,R$_{\odot}$, $\log T_{\rm eff}=3.875$ and $\log L/L_{\odot}=1.28$ and the binary-evolution model, showed in the right panel has the parameters: $M=2.1$\mass, $R=2.618$\,R$_{\odot}$, $\log T_{\rm eff}=3.886$ and $\log L/L_{\odot}=1.335$.

As can be seen the binary-evolution model shows instability in both low and intermediate frequency range whereas the single-evolution model is pulsational stable
in the whole range of frequencies. Thus Fig.\,\ref{opacity_OPAL} shows a direct effect of the binary evolution on the pulsational properties of $\delta$ Sct variables.
The conclusion is that an accreting matter has a huge impact on a star, not only in term of mass gain.
Therefore, computing the pulsations of $\delta$ Sct stars in binary systems such as KIC\,10661783 one cannot neglect a mass-exchange in the past.
Clearly, an incoming mass changes physical conditions and compositions of the outer layers. This fact is best demonstrated by the abundance profiles  of H and He
plotted in Fig.\,\ref{abundance_profiles}. Accumulation of helium in the outer layers of the primary has a great influence on excitation of pulsations.

\begin{figure*}
    \centering

    \begin{tabular}{cc}
    \multicolumn{2}{c}{\textbf{Modified OPAL opacities}} \\
    \textbf{Single star evolution} & \textbf{Binary evolution} \\
    \includegraphics[width=1.0\columnwidth,clip]{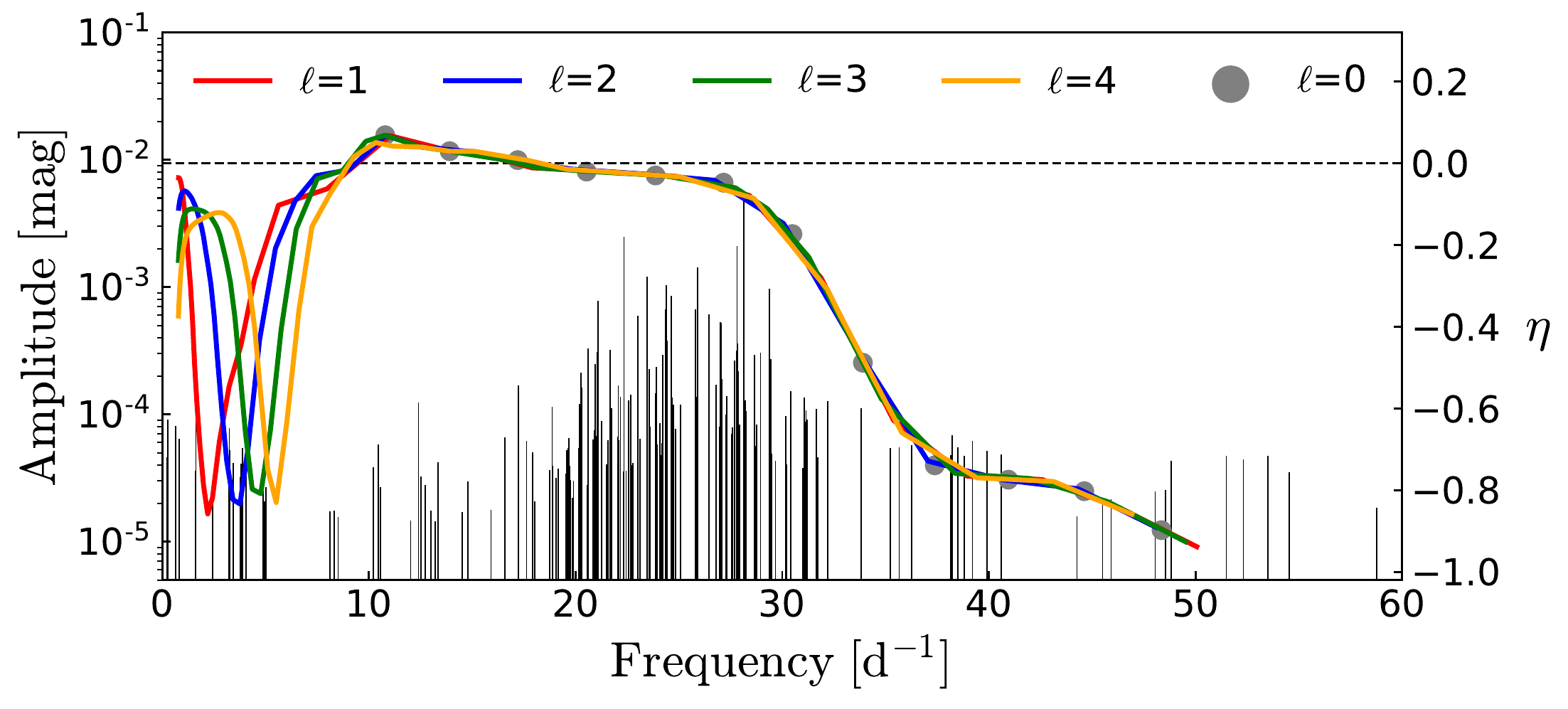}  &  \includegraphics[width=1.0\columnwidth,clip]{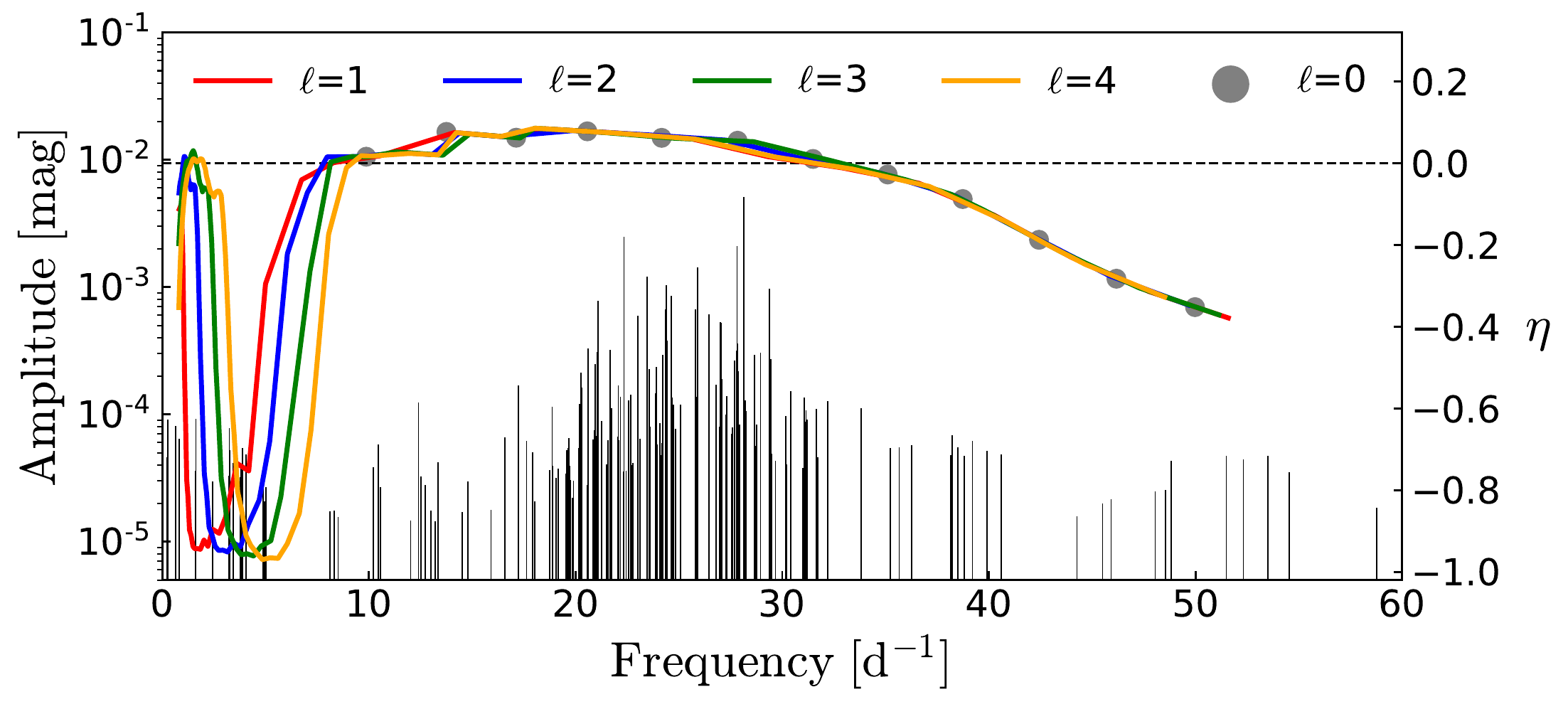}  \\
    \includegraphics[width=1.0\columnwidth,clip]{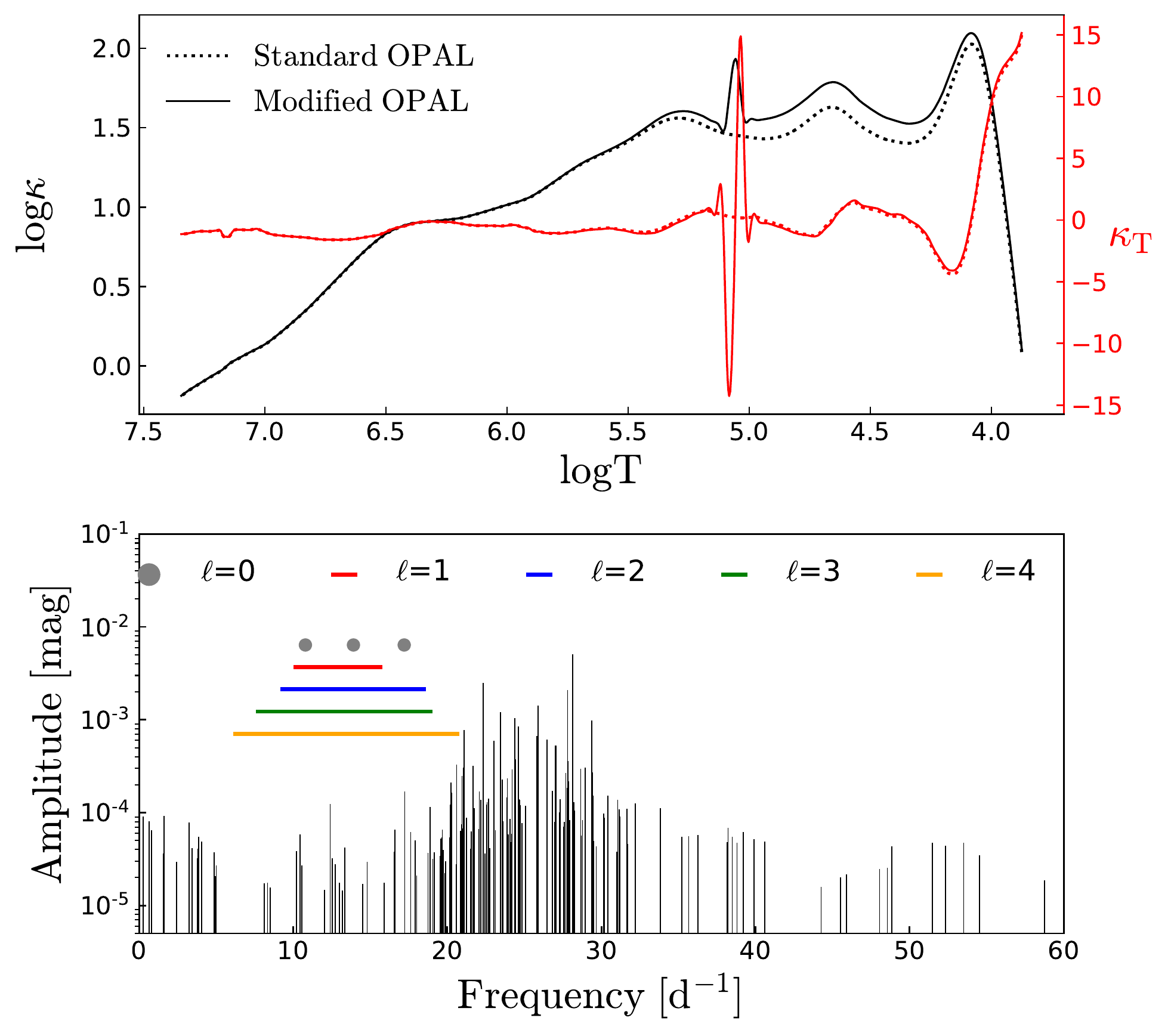}  &  \includegraphics[width=1.0\columnwidth,clip]{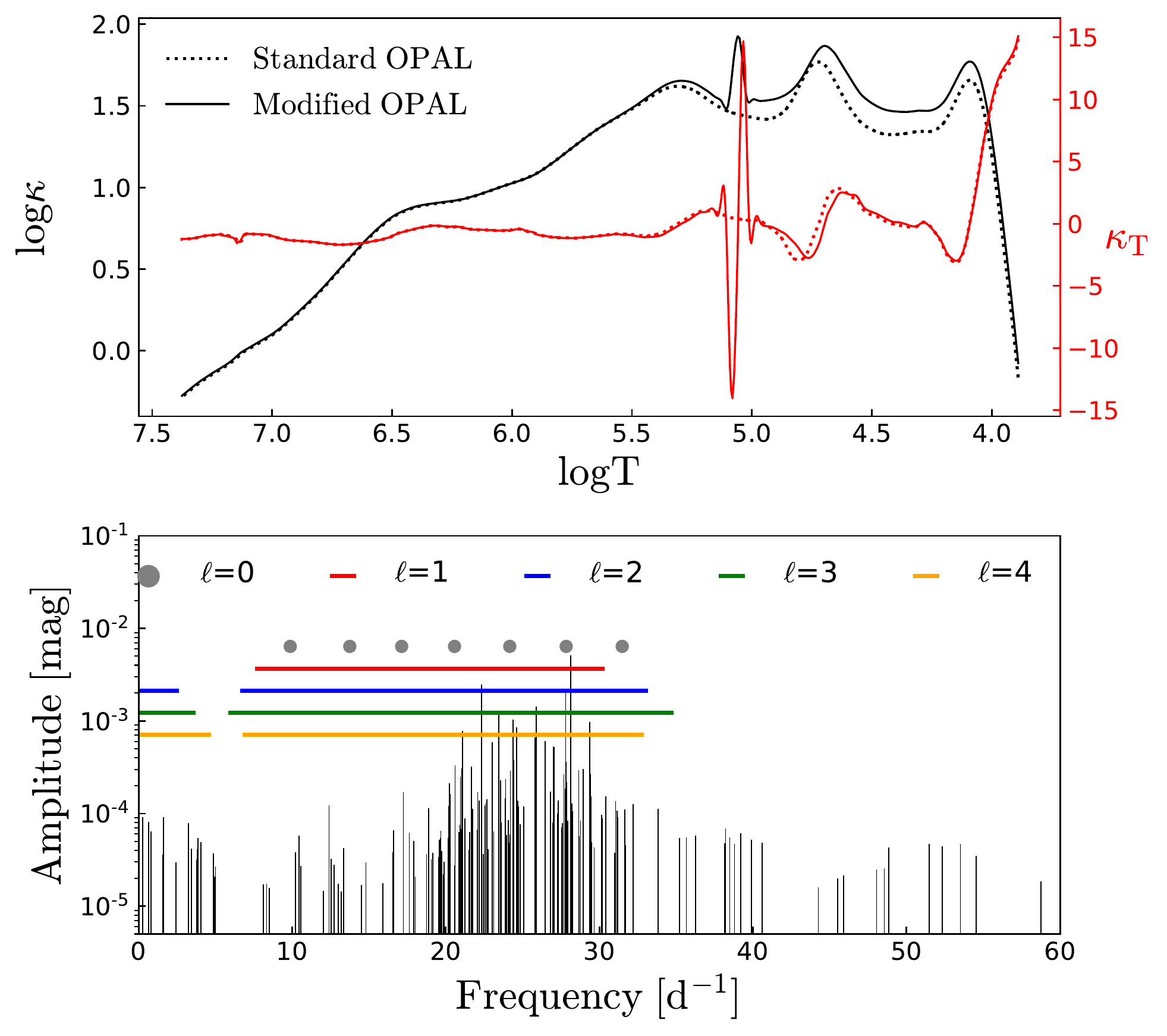}  \\
    \end{tabular}

    \caption{
    Top panels:  the instability parameter $\eta$ on the right Y-axis as a function of the frequency for modes with $\ell \le 4$ for single-evolution (left-hand side) and  binary-evolution (right-hand side) models. The left Y-axis gives the amplitudes of the observed frequencies. The representative models for single and binary case have very similar parameters (see text) and fit the observed values of $R$, $M$,
    $T_{\rm eff}$ and $\log L/L_{\sun}$ within their 3$\sigma$ errors. The models were calculated for $Z=0.025$, $X_0=0.73$, $f_{\rm ov}=0.02$, $\alpha_{\rm MLT}=1.3$ using the modified OPAL data with an increase of opacities by 100\% near $\log T=4.69$\,K and by 300\% near $\log T=5.06$\,K.
   The middle panels show the runs of the standard and modified mean opacities (black lines) and their logarithmic temperature derivatives (red lines). In bottom panels, we marked
   the range of rotationaly split unstable mode.}
    \label{opacity_mod}
\end{figure*}

In order to obtain the pulsational instability covering the whole observed frequency range, we followed the procedure of opacity modifications of \cite{Daszynska2017}. Since our binary model predicts low frequencies to be unstable, our main aim was to excite the modes with frequencies higher than 20\cpd. Because in $\delta$ Sct models, p modes are excited by the opacity mechanism acting in the partial HeII ionisation zone, we modified the OPAL tables by increasing the mean opacity near $\log T=4.69$\,K by 100\%.
Unfortunately, such modification decreased the instability of low frequency modes.
To prevent damping of gravity modes we increased the mean opacity at $\log T=5.06$\,K as suggested by \cite{Balona2015}. The increase by 300\% allowed us to obtain instabilities in both low and intermediate frequency ranges. We show that results in the top panels of Fig.\,\ref{opacity_mod} for both, single and binary evolution models, in a similar way as
in Fig.\,\ref{opacity_OPAL}. Moreover, the middle panels show a comparison of the standard and modified mean opacity profiles, $\kappa(T)$, as a function of temperature.
The logarithmic temperature derivatives, $\kappa_T = \partial \log \kappa (T) / \partial \log T$, are plotted with red lines.
We chose models matching, within 3$\sigma$, the observed values of the radius, effective temperature and luminosity. The single-evolution model has the parameters: $M=2.1$\mass, $R=2.582$\,R$_{\odot}$, $\log T_{\rm eff}=3.875$ and $\log L/L_{\odot}=1.278$ while the binary-evolution model has the parameters: $M=2.09$\mass, $R=2.545$\,R$_{\odot}$, $\log T_{\rm eff}=3.890$ and $\log L/L_{\odot}=1.325$.

Although we do not take into account the effects of rotation, at least rotational splitting of pulsational modes has to be included.
To this end, each theoretical frequency was split according to the formula
$$ f = f_0 + m f_{\rm rot}(1-C_{nl}),$$
where $f_0$ is the mode frequency in the non-rotating star, $m$ is the azimuthal order, $f_{\rm rot}$ is the rotational frequency of a star and $C_{nl}$ is a Ledoux constant dependent on a stellar structure and on the mode.
The primary of KIC\,10661783 rotates with the frequency of about 0.6\cpd, which is about 20\% of the critical value of the rotational frequency.
The ranges of the rotationally split unstable frequencies are marked in the bottom panels of Fig.\,\ref{opacity_mod}.
As one can see, by increasing the mean opacity near $\log T\approx4.69$ and $\log T\approx5.06$ and taking into account rotational splitting of pulsational modes, we are able to cover almost whole region of the observed frequencies, in particular those with the highest amplitudes. Unfortunately, the region of frequencies higher than 35\cpd remains stable no matter what modifications to the mean opacity profile we apply. We interpret this as a result of the convection treatment in the pulsational code we use,
that relies on the convective flux freezing approximation. Therefore, if the mechanism responsible for those mode excitation is based on the convection-pulsation interaction, as suggested by \cite{Antoci2014} and \cite{Xiong2016}, we cannot excite them in our models.

\section{Discussion and conclusions}
\label{sec:conclusions}

We performed complex studies of the binary system KIC\,10661783 with the main component being the pulsating star of $\delta$\,Sct type. Firstly, using the whole \textit{Kepler} photometry, we made the light curve modelling using the WD code. After subtraction of the eclipse light curve, we looked for variability in the residua by applying the Fourier analysis and standard pre-whitening procedure. This analysis allowed us to identify 590 significant frequencies, that is with S/N>4, with 207 of which being independent. Most of 207 peaks occupy the frequency range typical for $\delta$\,Sct pulsators but there are large number of low frequency peaks that, as in the case of other $\delta$ Scuti stars observed from the space, correspond most probably to high-order g mode pulsations.

Besides, we found a numerous orbital harmonics that can originate from the subtraction of imperfect eclipsing model or internal variability with the frequencies that are multiples of the orbital frequency. The last possibility can be associated with tidally excited pulsations. Moreover, the amplitude modulation with the orbital phase was found. Such behaviour is known from the tidally tilted pulsators and is interpreted as the variable pulsational amplitude over the stellar disk. An in-depth study of this phenomenon is beyond the scope of this paper.

We computed the binary evolution of the system using the \texttt{MESA} code that includes mass transfer as well as the evolution of orbital elements. We found the binary model that reproduces masses and radii of both components within the 3$\sigma$ error. That model also reproduces well the position of the components in the HR diagram, however the secondary extends slightly beyond the $3\sigma$ error box. It demanded to assume slightly non-conservative mass transfer with $\beta=0.05$. The other parameters of the best fit are: $X_0=0.73$, $Z=0.025$, $f_{\rm ov}=0.02$
and $\alpha_{\rm MLT}=1.3$. The internal structure of our model differs drastically from the single evolution one. In particular, due to the mass transfer episode, the outer layers of the main component are enormously enriched with helium and depleted of hydrogen. However, we cannot confront this result with observations because there is no determination of helium abundance in the literature. On the other hand, from our binary-evolution modelling, we obtained that abundances of CNO elements agree with the observational determinations within 2$\sigma$
except for the carbon of the primary.

Then, for the first time, we examined the impact of binary evolution, in term of the internal structure changes, on pulsational properties of $\delta$ Sct star models. We found that in the case of the single-evolution model, adequate for the main component, all pulsational modes are stable. Instead, the binary-evolution counterpart exhibits instability, both, for p modes and high-order g-modes. However, to cover the wider range of the observed frequencies the modification of opacity data was necessary. To this end, we increased the mean opacities by 100\% at the temperature $\log T=4.69$ (i.e., around the HeII ionization zone) and by 300\% at the temperature $\log T=5.06$. Including the rotational splitting of unstable modes allowed us to account for instability in the frequency range of about $(0,~35$\,d$^{-1}$). The frequencies higher than 35\,d$^{-1}$ are associated with high-order p modes excited, most probably, by another mechanism, e.g., the turbulent pressure in the H ionization zone, as proposed by \cite{Antoci2014}.

Our evolutionary and pulsational modelling clearly showed that such systems as KIC\,10661783 should be modelled as a binary and not as single stars as it is often done. These multi-faceted studies has led to the construction of a complex stellar model that explains the current stage of the binary system and both components, their evolutionary past, and account for pulsational instability in almost the entire range of the observed frequencies. The study of more binaries of this type may allow to draw more general conclusions on evolution and pulsation, and indicate the directions of development of asteroseismology of binary stars.

\section*{Acknowledgements}

This work was financially supported by the Polish National Science Centre grant 2018/29/B/ST9/02803.

Calculations have been carried out using resources provided by Wroc\l aw Centre for
Networking and Supercomputing (http://wcss.pl), grant no. 265.

Funding for the Kepler mission is provided by the NASA Science Mission directorate.
Some of the data presented in this paper were obtained from the Multi mission
Archive at the Space Telescope Science Institute (MAST). STScI is operated by the
Association of Universities for Research in Astronomy, Inc., under NASA contract.

\section*{Data Availability}

The target pixel files were downloaded from the public data archive at MAST. The light curves will be shared upon reasonable request. The full list of frequencies is available as a supplementary material to this paper. We make all inlists needed to recreate our \texttt{MESA-binary} results publicly available at Zenodo. Those can be downloaded at \url{https://doi.org/10.5281/zenodo.4618112}.




\bibliographystyle{mnras}
\interlinepenalty=10000
\bibliography{miszuda}



\appendix

\section{List of observed frequencies}
\label{sec:appendix}

\begin{table*}
    \caption{The table contains a full list of observed frequencies from SC data after comparing them with frequencies found in the LC data. For reference, $f_{\rm orb} = $ 0.8120027790\,\cpd.}
    \label{tab:freqs}
\begin{tabular}{lcccccccc}
        \hline
ID & Frequency $f(\rm d^{-1})$ & $\sigma_f (\rm d^{-1})$ & Amplitude $A(\rm mag)$ & $\sigma_A (\rm mag)$ & Phase $\Phi (\rm rad)$ & $\sigma_{\Phi} (\rm rad)$ & S/N & Remarks \\
        \hline
        \hline
$f_{ 1 } $ & $  28.1344873410 $ & $   0.0000013012 $ & $   0.0050862332 $ & $   0.0000031072 $ & $   0.4657728292 $ & $   0.0000971534 $ & $  42.57 $ & \\
$f_{ 2 } $ & $  22.3405933680 $ & $   0.0000028559 $ & $   0.0024871517 $ & $   0.0000074126 $ & $   0.5238758669 $ & $   0.0004740919 $ & $  44.20 $ & \\
$f_{ 3 } $ & $  27.8105546960 $ & $   0.0000027177 $ & $   0.0020998242 $ & $   0.0000024453 $ & $   0.3964413422 $ & $   0.0001852580 $ & $  36.00 $ & \\
$f_{ 4 } $ & $  25.9032184080 $ & $   0.0000041336 $ & $   0.0014309774 $ & $   0.0000025583 $ & $   0.8405118105 $ & $   0.0002846019 $ & $  28.90 $ & \\
$f_{ 5 } $ & $  29.7586885110 $ & $   0.0000042774 $ & $   0.0012120314 $ & $   0.0000027064 $ & $   0.7144856101 $ & $   0.0003551001 $ & $  25.31 $ & $f_{    5} +     2f_{\rm orb} $ \\
$f_{ 6 } $ & $  23.4611786120 $ & $   0.0000041648 $ & $   0.0012072406 $ & $   0.0000025275 $ & $   0.6234627829 $ & $   0.0003333014 $ & $  30.91 $ & \\
$f_{ 7 } $ & $  24.4215255930 $ & $   0.0000047223 $ & $   0.0009883436 $ & $   0.0000024078 $ & $   0.2008024590 $ & $   0.0003877161 $ & $  18.92 $ & $   10f_{    4}   -10f_{    6} $ \\
\hline
\end{tabular}
\end{table*}

\begin{table*}
  \contcaption{}
    \begin{tabular}{lcccccccc}
        \hline
        ID & Frequency $f(\rm d^{-1})$ & $\sigma_f (\rm d^{-1})$ & Amplitude $A(\rm mag)$ & $\sigma_A (\rm mag)$ & Phase $\Phi (\rm rad)$ & $\sigma_{\Phi} (\rm rad)$ & S/N & Remarks \\
        \hline
        \hline
$f_{ 71 } $ & $  22.5789374680 $ & $   0.0000155792 $ & $   0.0001289062 $ & $   0.0000021732 $ & $   0.9758592986 $ & $   0.0026842049 $ & $  10.67 $ & \\
$f_{ 72 } $ & $  30.4065238770 $ & $   0.0000160455 $ & $   0.0001532298 $ & $   0.0000022407 $ & $   0.6024479714 $ & $   0.0023285458 $ & $  12.00 $ & \\
$f_{ 73 } $ & $  20.2890469820 $ & $   0.0000162300 $ & $   0.0001632853 $ & $   0.0000020320 $ & $   0.1463468113 $ & $   0.0019805174 $ & $  13.27 $ & \\
$f_{ 74 } $ & $  25.8121233670 $ & $   0.0000168848 $ & $   0.0001602849 $ & $   0.0000020499 $ & $   0.8162558425 $ & $   0.0020354168 $ & $   6.33 $ & \\
$f_{ 75 } $ & $  27.7940389720 $ & $   0.0000167202 $ & $   0.0001851723 $ & $   0.0000020702 $ & $   0.1182624675 $ & $   0.0017800195 $ & $   6.36 $ & \\
$f_{ 76 } $ & $  29.4344299650 $ & $   0.0000145760 $ & $   0.0002435661 $ & $   0.0000024097 $ & $   0.6562810709 $ & $   0.0015751429 $ & $   9.58 $ & $f_{   76} +     3f_{\rm orb} $ \\
$f_{ 77 } $ & $  25.4438901330 $ & $   0.0000167077 $ & $   0.0001332128 $ & $   0.0000020578 $ & $   0.3564789455 $ & $   0.0024589495 $ & $   6.16 $ & $    7f_{   55}    -5f_{   74} $ \\
\hline
\end{tabular}
\end{table*}

\begin{table*}
  \contcaption{}
    \begin{tabular}{lcccccccc}
        \hline
        ID & Frequency $f(\rm d^{-1})$ & $\sigma_f (\rm d^{-1})$ & Amplitude $A(\rm mag)$ & $\sigma_A (\rm mag)$ & Phase $\Phi (\rm rad)$ & $\sigma_{\Phi} (\rm rad)$ & S/N & Remarks \\
        \hline
        \hline
$f_{ 151 } $ & $  24.6305928140 $ & $   0.0000244636 $ & $   0.0001102442 $ & $   0.0000020814 $ & $   0.2398382048 $ & $   0.0030049243 $ & $   5.63 $ & $f_{  151}    -f_{\rm orb} $ \\
$f_{ 152 } $ & $  55.9450332450 $ & $   0.0000232863 $ & $   0.0000934078 $ & $   0.0000020126 $ & $   0.3186026993 $ & $   0.0034293117 $ & $  11.30 $ & $    2f_{  152}    -6f_{\rm orb} $ \\
$f_{ 155 } $ & $  23.3759309440 $ & $   0.0000233581 $ & $   0.0000918441 $ & $   0.0000020183 $ & $   0.6550578723 $ & $   0.0034960959 $ & $   6.97 $ & $f_{  155}    -3f_{\rm orb} $ \\
$f_{ 156 } $ & $  27.0313653980 $ & $   0.0000242017 $ & $   0.0001102353 $ & $   0.0000021064 $ & $   0.9440311032 $ & $   0.0030412942 $ & $   4.62 $ & \\
$f_{ 157 } $ & $  27.0415391700 $ & $   0.0000242223 $ & $   0.0001050919 $ & $   0.0000020672 $ & $   0.2148550551 $ & $   0.0031311255 $ & $   4.68 $ & \\
$f_{ 158 } $ & $  33.0071436350 $ & $   0.0000234463 $ & $   0.0000989165 $ & $   0.0000020274 $ & $   0.2332416605 $ & $   0.0032647545 $ & $  10.97 $ & $f_{  158}    -f_{\rm orb} $ \\
$f_{ 159 } $ & $  24.5743007700 $ & $   0.0000235916 $ & $   0.0000942113 $ & $   0.0000020215 $ & $   0.4630808145 $ & $   0.0034162343 $ & $   5.63 $ & $f_{  159}    -4f_{\rm orb} $ \\
\hline
\end{tabular}
\end{table*}

\begin{table*}
  \contcaption{}
    \begin{tabular}{lcccccccc}
        \hline
        ID & Frequency $f(\rm d^{-1})$ & $\sigma_f (\rm d^{-1})$ & Amplitude $A(\rm mag)$ & $\sigma_A (\rm mag)$ & Phase $\Phi (\rm rad)$ & $\sigma_{\Phi} (\rm rad)$ & S/N & Remarks \\
        \hline
        \hline
$f_{ 236 } $ & $  24.8282615500 $ & $   0.0000300434 $ & $   0.0000675676 $ & $   0.0000020232 $ & $   0.9331451078 $ & $   0.0047677038 $ & $   4.22 $ & \\
$f_{ 238 } $ & $  28.5313206400 $ & $   0.0000297991 $ & $   0.0000676273 $ & $   0.0000020318 $ & $   0.7839837476 $ & $   0.0047796520 $ & $   4.52 $ & $    2f_{   75}    -f_{   65} $ \\
$f_{ 239 } $ & $  24.9868835440 $ & $   0.0000301999 $ & $   0.0000654944 $ & $   0.0000020161 $ & $   0.8471002988 $ & $   0.0048976301 $ & $   4.15 $ & $   10f_{  157}    -9f_{  181} $ \\
$f_{ 241 } $ & $  24.6369975280 $ & $   0.0000329121 $ & $   0.0000708569 $ & $   0.0000020902 $ & $   0.0503229238 $ & $   0.0046931449 $ & $   4.45 $ & $f_{  135} +    10f_{  187} $ \\
$f_{ 241 } $ & $  24.6369975280 $ & $   0.0000329121 $ & $   0.0000708569 $ & $   0.0000020902 $ & $   0.0503229238 $ & $   0.0046931449 $ & $   4.45 $ & $f_{  135} +     9f_{  187} $ \\
$f_{ 242 } $ & $  21.7692582990 $ & $   0.0000303436 $ & $   0.0000639935 $ & $   0.0000020227 $ & $   0.2857181609 $ & $   0.0050312728 $ & $   6.48 $ & $f_{  242} + f_{\rm orb} $ \\
$f_{ 243 } $ & $  25.8710382240 $ & $   0.0000292151 $ & $   0.0000688516 $ & $   0.0000020656 $ & $   0.6194239018 $ & $   0.0047754854 $ & $   4.06 $ & $    2f_{  243}    -9f_{\rm orb} $ \\
\hline
\end{tabular}
\end{table*}

\begin{table*}
  \contcaption{}
    \begin{tabular}{lcccccccc}
        \hline
        ID & Frequency $f(\rm d^{-1})$ & $\sigma_f (\rm d^{-1})$ & Amplitude $A(\rm mag)$ & $\sigma_A (\rm mag)$ & Phase $\Phi (\rm rad)$ & $\sigma_{\Phi} (\rm rad)$ & S/N & Remarks \\
        \hline
        \hline
$f_{ 316 } $ & $  34.0356694400 $ & $   0.0000360788 $ & $   0.0000515384 $ & $   0.0000020169 $ & $   0.0905508861 $ & $   0.0062258598 $ & $   5.73 $ & $f_{  316}    -2f_{\rm orb} $ \\
$f_{ 317 } $ & $   2.4339260800 $ & $   0.0000315934 $ & $   0.0000776335 $ & $   0.0000023228 $ & $   0.9747783557 $ & $   0.0047139259 $ & $   7.88 $ & $f_{  317}    -f_{\rm orb} $ \\
$f_{ 319 } $ & $  17.9861244730 $ & $   0.0000364713 $ & $   0.0000539482 $ & $   0.0000020193 $ & $   0.3442906320 $ & $   0.0059572648 $ & $   7.78 $ & $f_{  319}    -5f_{\rm orb} $ \\
$f_{ 320 } $ & $  22.0759969820 $ & $   0.0000465285 $ & $   0.0000626241 $ & $   0.0000021666 $ & $   0.8593286980 $ & $   0.0055076165 $ & $   5.07 $ & \\
$f_{ 321 } $ & $  23.2128085380 $ & $   0.0000364120 $ & $   0.0000534370 $ & $   0.0000020217 $ & $   0.6894532886 $ & $   0.0060196388 $ & $   4.98 $ & $    6f_{   19}    -5f_{   27} $ \\
$f_{ 322 } $ & $  52.3958806960 $ & $   0.0000367646 $ & $   0.0000507997 $ & $   0.0000020127 $ & $   0.0725169593 $ & $   0.0063065674 $ & $   4.50 $ & $    4f_{  322}    -7f_{\rm orb} $ \\
$f_{ 323 } $ & $  24.3629832980 $ & $   0.0000357104 $ & $   0.0000600872 $ & $   0.0000022157 $ & $   0.5284895695 $ & $   0.0058695406 $ & $   4.02 $ & $   30f_{\rm orb} $ \\
\hline
\end{tabular}
\end{table*}

\begin{table*}
  \contcaption{}
    \begin{tabular}{lcccccccc}
        \hline
        ID & Frequency $f(\rm d^{-1})$ & $\sigma_f (\rm d^{-1})$ & Amplitude $A(\rm mag)$ & $\sigma_A (\rm mag)$ & Phase $\Phi (\rm rad)$ & $\sigma_{\Phi} (\rm rad)$ & S/N & Remarks \\
        \hline
        \hline
$f_{ 396 } $ & $   3.4461557650 $ & $   0.0000437917 $ & $   0.0000416814 $ & $   0.0000020142 $ & $   0.4530661716 $ & $   0.0076868116 $ & $   5.03 $ & \\
$f_{ 398 } $ & $  30.1704546840 $ & $   0.0000414810 $ & $   0.0000558340 $ & $   0.0000022913 $ & $   0.9023431481 $ & $   0.0065321596 $ & $   4.18 $ & \\
$f_{ 399 } $ & $  30.8980978550 $ & $   0.0000438771 $ & $   0.0000433983 $ & $   0.0000020179 $ & $   0.2112838224 $ & $   0.0073994937 $ & $   4.29 $ & $f_{  399}    -f_{\rm orb} $ \\
$f_{ 402 } $ & $   3.8086755260 $ & $   0.0000441123 $ & $   0.0000409501 $ & $   0.0000020232 $ & $   0.0444221631 $ & $   0.0078648524 $ & $   5.07 $ & \\
$f_{ 403 } $ & $   3.8229697580 $ & $   0.0000439120 $ & $   0.0000418743 $ & $   0.0000020222 $ & $   0.2537245669 $ & $   0.0076856760 $ & $   5.20 $ & $f_{   28}    -f_{   30} $ \\
$f_{ 404 } $ & $  32.1455181680 $ & $   0.0000439136 $ & $   0.0000416311 $ & $   0.0000020178 $ & $   0.0477770622 $ & $   0.0077144495 $ & $   4.89 $ & $    2f_{  123}    -f_{  146} $ \\
$f_{ 405 } $ & $  55.2737771200 $ & $   0.0000440846 $ & $   0.0000411026 $ & $   0.0000020138 $ & $   0.6792965616 $ & $   0.0078010477 $ & $   5.15 $ & $    9f_{  405}    -4f_{\rm orb} $ \\
\hline
\end{tabular}
\end{table*}

\begin{table*}
  \contcaption{}
    \begin{tabular}{lcccccccc}
        \hline
        ID & Frequency $f(\rm d^{-1})$ & $\sigma_f (\rm d^{-1})$ & Amplitude $A(\rm mag)$ & $\sigma_A (\rm mag)$ & Phase $\Phi (\rm rad)$ & $\sigma_{\Phi} (\rm rad)$ & S/N & Remarks \\
        \hline
        \hline
$f_{ 485 } $ & $  47.3465266090 $ & $   0.0000536549 $ & $   0.0000330867 $ & $   0.0000020117 $ & $   0.6353809935 $ & $   0.0096803158 $ & $   6.17 $ & $    2f_{  221}    -4f_{  285} $ \\
$f_{ 487 } $ & $  19.2247844610 $ & $   0.0000538821 $ & $   0.0000305744 $ & $   0.0000020198 $ & $   0.4030146274 $ & $   0.0105125821 $ & $   4.93 $ & $f_{  487}    -2f_{\rm orb} $ \\
$f_{ 488 } $ & $  17.3767390980 $ & $   0.0000536477 $ & $   0.0000376929 $ & $   0.0000020395 $ & $   0.5967206171 $ & $   0.0086103933 $ & $   5.72 $ & $f_{  488} + f_{\rm orb} $ \\
$f_{ 489 } $ & $  16.5646760160 $ & $   0.0000537398 $ & $   0.0000379836 $ & $   0.0000020345 $ & $   0.1852448903 $ & $   0.0085233969 $ & $   6.04 $ & \\
$f_{ 490 } $ & $  32.2961972130 $ & $   0.0000541582 $ & $   0.0000315537 $ & $   0.0000020150 $ & $   0.3322494670 $ & $   0.0101600460 $ & $   4.39 $ & $f_{  505} +     3f_{  502} $ \\
$f_{ 491 } $ & $  16.3005563770 $ & $   0.0000542079 $ & $   0.0000351089 $ & $   0.0000020256 $ & $   0.9108406108 $ & $   0.0091794766 $ & $   6.95 $ & $f_{  491}    -f_{\rm orb} $ \\
$f_{ 492 } $ & $  19.5489811980 $ & $   0.0000437837 $ & $   0.0000490668 $ & $   0.0000023056 $ & $   0.4048889847 $ & $   0.0074745801 $ & $   5.14 $ & $f_{  492} +     2f_{\rm orb} $ \\
\hline
\end{tabular}
\end{table*}

\begin{table*}
  \contcaption{}
    \begin{tabular}{lcccccccc}
        \hline
        ID & Frequency $f(\rm d^{-1})$ & $\sigma_f (\rm d^{-1})$ & Amplitude $A(\rm mag)$ & $\sigma_A (\rm mag)$ & Phase $\Phi (\rm rad)$ & $\sigma_{\Phi} (\rm rad)$ & S/N & Remarks \\
        \hline
        \hline
$f_{ 578 } $ & $  19.1068648610 $ & $   0.0000695605 $ & $   0.0000251862 $ & $   0.0000020219 $ & $   0.1384747357 $ & $   0.0127737495 $ & $   4.34 $ & $    7f_{  278}   -10f_{  374} $ \\
$f_{ 579 } $ & $  48.0482018540 $ & $   0.0000695910 $ & $   0.0000248175 $ & $   0.0000020124 $ & $   0.9841352193 $ & $   0.0129049974 $ & $   4.25 $ & \\
$f_{ 580 } $ & $  19.8719948700 $ & $   0.0000695073 $ & $   0.0000236512 $ & $   0.0000020277 $ & $   0.3222179430 $ & $   0.0136500928 $ & $   4.33 $ & $f_{  580}    -2f_{\rm orb} $ \\
$f_{ 581 } $ & $  12.0770707070 $ & $   0.0000711531 $ & $   0.0000223215 $ & $   0.0000020311 $ & $   0.9429091484 $ & $   0.0144881019 $ & $   6.08 $ & $    5f_{  600}    -6f_{  648} $ \\
$f_{ 582 } $ & $  11.7175513320 $ & $   0.0000710902 $ & $   0.0000244705 $ & $   0.0000020197 $ & $   0.9206722549 $ & $   0.0131413176 $ & $   6.01 $ & $f_{  582}    -f_{\rm orb} $ \\
$f_{ 585 } $ & $  18.6785867290 $ & $   0.0000715461 $ & $   0.0000242179 $ & $   0.0000020572 $ & $   0.6526188960 $ & $   0.0135069982 $ & $   4.23 $ & $   23f_{\rm orb} $ \\
$f_{ 586 } $ & $  44.3210436920 $ & $   0.0000728895 $ & $   0.0000233745 $ & $   0.0000020124 $ & $   0.8575422625 $ & $   0.0136995638 $ & $   5.84 $ & $    6f_{  250}    -9f_{  647} $ \\
\hline
\end{tabular}
\end{table*}

\begin{table*}
  \contcaption{}
    \begin{tabular}{lcccccccc}
        \hline
        ID & Frequency $f(\rm d^{-1})$ & $\sigma_f (\rm d^{-1})$ & Amplitude $A(\rm mag)$ & $\sigma_A (\rm mag)$ & Phase $\Phi (\rm rad)$ & $\sigma_{\Phi} (\rm rad)$ & S/N & Remarks \\
        \hline
        \hline
$f_{ 670 } $ & $  13.2003127180 $ & $   0.0001161685 $ & $   0.0000144908 $ & $   0.0000020130 $ & $   0.3094334100 $ & $   0.0221160418 $ & $   4.29 $ & \\
$f_{ 671 } $ & $   9.5015820780 $ & $   0.0001193975 $ & $   0.0000142210 $ & $   0.0000020130 $ & $   0.8883329073 $ & $   0.0225197344 $ & $   4.20 $ & $    4f_{   71}    -4f_{  111} $ \\
$f_{ 672 } $ & $  15.5150565210 $ & $   0.0001206529 $ & $   0.0000141865 $ & $   0.0000020151 $ & $   0.6633033484 $ & $   0.0226106309 $ & $   4.05 $ & $    2f_{  553}   -10f_{  549} $ \\
$f_{ 673 } $ & $  12.0695782580 $ & $   0.0001204656 $ & $   0.0000143094 $ & $   0.0000020353 $ & $   0.2295614778 $ & $   0.0226251286 $ & $   4.13 $ & $f_{  673} +     2f_{\rm orb} $ \\
$f_{ 675 } $ & $  15.4713740070 $ & $   0.0001255827 $ & $   0.0000138216 $ & $   0.0000020140 $ & $   0.0122204946 $ & $   0.0231731866 $ & $   4.07 $ & $    8f_{  654}    -2f_{  660} $ \\
$f_{ 676 } $ & $  86.0834542500 $ & $   0.0001301903 $ & $   0.0000143129 $ & $   0.0000020220 $ & $   0.0230234552 $ & $   0.0224865277 $ & $   7.71 $ & $  106f_{\rm orb} $ \\
$f_{ 677 } $ & $  86.8981785640 $ & $   0.0001326842 $ & $   0.0000136480 $ & $   0.0000021966 $ & $   0.8811951073 $ & $   0.0256151333 $ & $   8.16 $ & $    3f_{  677} + f_{\rm orb} $ \\
\hline
\end{tabular}
\end{table*}

\begin{table*}
  \contcaption{}
    \begin{tabular}{lcccccccc}
        \hline
        ID & Frequency $f(\rm d^{-1})$ & $\sigma_f (\rm d^{-1})$ & Amplitude $A(\rm mag)$ & $\sigma_A (\rm mag)$ & Phase $\Phi (\rm rad)$ & $\sigma_{\Phi} (\rm rad)$ & S/N & Remarks \\
        \hline
        \hline
$f_{ 737 } $ & $ 172.1666802340 $ & $   0.0003939587 $ & $   0.0000045405 $ & $   0.0000020189 $ & $   0.4471206978 $ & $   0.0707483643 $ & $   5.05 $ & $  212f_{\rm orb} $ \\
$f_{ 738 } $ & $ 168.9182866860 $ & $   0.0003879242 $ & $   0.0000043808 $ & $   0.0000020178 $ & $   0.9220538368 $ & $   0.0733195946 $ & $   4.77 $ & $  208f_{\rm orb} $ \\
$f_{ 739 } $ & $ 143.7430813000 $ & $   0.0004054032 $ & $   0.0000046529 $ & $   0.0000020210 $ & $   0.8371635562 $ & $   0.0691508510 $ & $   4.57 $ & $  177f_{\rm orb} $ \\
$f_{ 740 } $ & $ 145.3673102240 $ & $   0.0003917032 $ & $   0.0000046043 $ & $   0.0000020211 $ & $   0.5818380113 $ & $   0.0698736048 $ & $   4.61 $ & $  179f_{\rm orb} $ \\
$f_{ 741 } $ & $ 168.1061983140 $ & $   0.0004107385 $ & $   0.0000041550 $ & $   0.0000020187 $ & $   0.5159911230 $ & $   0.0773226105 $ & $   4.61 $ & $  207f_{\rm orb} $ \\
$f_{ 742 } $ & $ 160.7974660900 $ & $   0.0004169049 $ & $   0.0000041211 $ & $   0.0000020157 $ & $   0.5504041081 $ & $   0.0778729289 $ & $   4.67 $ & $  198f_{\rm orb} $ \\
$f_{ 743 } $ & $ 123.4405576810 $ & $   0.0004173544 $ & $   0.0000040914 $ & $   0.0000020196 $ & $   0.4949585859 $ & $   0.0785534446 $ & $   4.03 $ & $  152f_{\rm orb} $ \\
\hline
\end{tabular}
\end{table*}


\bsp    
\label{lastpage}
\end{document}